\documentclass{article}
\usepackage[onecolumn]{emulateapj}
\usepackage{epsf}

\def\lae{\mathrel{<\kern-1.0em\lower0.9ex\hbox{$\sim$}}}
\def\gae{\mathrel{>\kern-1.0em\lower0.9ex\hbox{$\sim$}}}
 
\begin{document}
 
\submitted{Accepted for Publication in the Astrophysical Journal}

\title{Dynamics of the Globular Cluster System Associated with M87 (NGC 4486). 
II. Analysis}
\author{Patrick C\^ot\'e\altaffilmark{1,2,3,4}, Dean E.
McLaughlin\altaffilmark{4,5,6}, David A. Hanes\altaffilmark{4,7,8,14}, Terry J.
Bridges\altaffilmark{4,8,9,10}, Doug Geisler\altaffilmark{11,12,13}, David
Merritt\altaffilmark{2}, James E. Hesser\altaffilmark{14}, 
Gretchen L.H. Harris\altaffilmark{15}, Myung Gyoon Lee\altaffilmark{16}}

\altaffiltext{1}{California Institute of Technology, Mail Stop 105-24,
Pasadena, CA 91125, USA}

\altaffiltext{2}{Department of Physics and Astronomy, Rutgers University,
New Brunswick, NJ 08854, USA}

\altaffiltext{3}{Sherman M. Fairchild Fellow}

\altaffiltext{4}{Visiting Astronomer, Canada-France-Hawaii Telescope,
operated by the National Research  Council of Canada, the Centre National
de la Recherche Scientifique of France, and the University of Hawaii.}

\altaffiltext{5}{Department of Astronomy, 601 Campbell Hall, University
of California, Berkeley, CA 94720-3411, USA}

\altaffiltext{6}{Hubble Fellow}

\altaffiltext{7}{Department of Physics, Queen's University, Kingston,
ON K7L 3N6, Canada}

\altaffiltext{8}{Anglo-Australian Observatory, P.O. Box 296, Epping, NSW,
1710, Australia}

\altaffiltext{9}{Royal Greenwich Observatory, Madingley Road, Cambridge,
CB3 0EZ, UK}

\altaffiltext{10}{Institute of Astronomy, Madingley Road, Cambridge,
CB3 0HA, UK}

\altaffiltext{11}{Grupo de Astronom\a'{\i}a
Dpto. de F\a'{\i}sica, Universidad de Concepci\'on,
Casilla 160-C, Concepci\'on, Chile}

\altaffiltext{12}{Visiting Astronomer, Cerro Tololo Inter-American Observatory,
which is operated by AURA, Inc., under cooperative agreement with the National
Science Foundation.}

\altaffiltext{13}{Visiting Astronomer, Kitt Peak National Observatory, which is
operated by AURA, Inc., under cooperative agreement with the National Science
Foundation.}

\altaffiltext{14}{Dominion Astrophysical Observatory, Herzberg Institute of
Astrophysics, National Research Council, 5071 West Saanich Road, Victoria, BC
V8W 3P6, Canada}

\altaffiltext{15}{Department of Physics, University of Waterloo, Waterloo ON
N2L 3G1, Canada}

\altaffiltext{16}{Astronomy Program, SEES, Seoul National University, Seoul
151-742, Korea}

\lefthead{{\sc C\^OT\'E ET AL.}}
\righthead{DYNAMICS OF THE M87 GLOBULAR CLUSTER SYSTEM}

\begin{abstract}

We present a dynamical analysis of the globular cluster system associated with
M87 (= NGC 4486), the cD galaxy near the dynamical center of
the Virgo cluster. The analysis utilizes a new spectroscopic and photometric 
database which is described in a companion paper (Hanes et al. 2001). Using a sample 
of 278 globular clusters with measured radial velocities and metallicities, and new 
surface density profiles based on wide-field Washington photometry, we 
study the dynamics of the M87 globular cluster system both globally --- for the entire 
cluster sample --- and separately --- for the metal-rich 
and metal-poor globular cluster samples. This constitutes the largest 
sample of radial velocities for pure Population II tracers yet assembled for
any galaxy. Our principal findings are summarized as follows: 

\begin{itemize}

\item[(1)] Surface density profiles constructed from our Washington photometry reveal
the metal-poor cluster system to be more spatially extended than its metal-rich 
counterpart, consistent with earlier findings based on {\sl HST} imaging in the central
regions of the galaxy.
Beyond a radius of $R \simeq 1.5R_e$ (10 kpc), the metal-poor component 
dominates the total globular cluster system.

\item[(2)] When considered in their entirety, each of the combined, metal-poor and metal-rich 
globular cluster samples (278, 161 and 117 clusters, respectively) appear to rotate, with 
similar rotation amplitudes, about axes whose position angles are indistinguishable from that 
of the photometric minor axis, $\Theta_0 = 65^{\circ}$.

\item[(3)] The one-dimensional rotation curve (i.e., binned in circular annuli) for the 
metal-rich cluster system has a roughly constant mean amplitude of 
$\Omega R = 160^{+120}_{-99}$ km s$^{-1}$.
The metal-rich clusters appear to be rotating, at all radii, about the photometric
minor axis of the galaxy. However, a smoothed, two-dimensional map of the line-of-sight
velocity residuals suggests that the rotation field for the metal-rich clusters is
non-cylindrical in nature. Instead, it exhibits a ``double-lobed" pattern, with maxima at 
$R \sim$ 3.5-4$R_e$ (25-30 kpc) along the approximate photometric major axis of the galaxy.

\item[(4)] The one-dimensional rotation curve of the metal-poor 
cluster system has mean amplitude of $\Omega R = 172^{+51}_{-108}$ km s$^{-1}$. The
two-dimensional map of the rotation field for the metal-poor clusters shows some
evidence for solid-body rotation or, alternatively, for a ``shear" in the line-of-sight velocity.
This shear is similar in size and orientation to that observed for Virgo {\it galaxies} 
within two degrees of M87, and is consistent with a scenario, 
previously suggested on the basis of dwarf galaxy kinematics and x-ray imaging, in which 
material is gradually infalling onto M87 along the so-called ``principal axis" 
of the Virgo cluster.

\item[(5)] Beyond a radius of $R \simeq 2R_e$ (15 kpc), the approximate onset of the galaxy's
cD envelope, the metal-poor globular cluster system rotates about the 
photometric minor axis --- similar to its metal-rich counterpart. Inside this
radius, however, the metal-poor clusters appear to rotate around the
photometric {\sl major} axis.

\item[(6)] The complete sample of 278 globular clusters has an almost perfectly isotropic
velocity ellipsoid, with $\beta_{\rm cl} = 1 - {\sigma_{\theta}^2 / {\sigma}_r^2} \simeq 0$.

\item[(7)] When considered separately, the metal-poor cluster system shows a modest
but significant {\it tangential} bias of 
$\beta_{\rm cl} \simeq -0.4$, 
while the velocity ellipsoid of the metal-rich cluster system is {\it radially} 
biased, with $\beta_{\rm cl} \simeq +0.4$. 

\end{itemize}

Taken together, these results demonstrate that the dual nature of the M87 
globular cluster system --- first identified on the basis of its bimodal metallicity 
distribution --- also extends to its dynamical properties. We discuss the implications 
of these findings for the various formation scenarios proposed for giant elliptical 
galaxies and their globular cluster systems.  \end{abstract}
 
 
\section{Introduction}

As surviving relics from the epoch of galaxy formation, globular clusters (GCs) 
provide important insights into the processes which shaped their present day 
host galaxies. The observed properties of GCs in giant elliptical galaxies --- 
such as their total numbers, chemical abundances and spatial distributions --- 
have been used alternatively to argue for a wide 
diversity of formation processes including dissipative (monolithic) collapse,
spiral-spiral mergers (Ashman \& Zepf 1992), multi-modal 
star formation histories caused by galactic winds (Harris, Harris \& 
McLaughlin 1998), and dissipationless hierarchical growth 
(C\^ot\'e, Marzke \& West 1998; C\^ot\'e et al. 2000). There is, however, one 
important piece of observational evidence which has been almost entirely absent 
in such discussions --- the {\it dynamical properties} of 
the GCs. Due to the faintness of even the most luminous GCs
in nearby giant elliptical galaxies, the requisite spectroscopic 
observations are extremely challenging, and it is only recently that radial 
velocities for significant samples of clusters have been accumulated 
in a handful of galaxies (e.g., Cohen \& Ryzhov 1997; Sharples et~al. 1998; 
Zepf et~al. 2000).

M87, the incipient cD galaxy near the dynamical center
of the Virgo cluster, offers an ideal testing ground for the various formation
scenarios. It has by far the most populous globular cluster system (GCS) of any 
nearby galaxy (e.g., N$_{\rm cl}$ $\simeq$ 13,500; McLaughlin et al. 1994), 
and a wealth of observational data has been accumulated for both the 
galaxy itself, and the surrounding Virgo cluster. 
Such material includes numerous analyses of the photometric properties and 
spatial distribution of M87 GCs ($e.g$., Strom et al. 1981; 
McLaughlin et al. 1994; Geisler, Lee \& Kim 2001), surface brightness profiles 
for the galaxy light (de Vaucouleurs \& Nieto 1978; Carter \& Dixon 1978), 
stellar dynamical studies of the central regions of M87 (van der Marel 1994),
photometric and radial velocity surveys of surrounding Virgo cluster 
galaxies (Binggeli, Sandage \& Tammann 1985; Binggeli, Popescu \& Tammann 1993;
Girardi et al. 1996), and X-ray observations of the hot intracluster gas
which fills the Virgo gravitational potential well (Nulsen \& B\"ohringer
1995).

The dynamics of the M87 GCS have been addressed in several previous studies, 
albeit in varying levels of detail. Early efforts (Huchra \& Brodie 1987; 
Mould, Oke \& Nemec 1987; Mould et al. 1990) were hampered by limited sample 
sizes: for example, the compilation of Mould et al. (1990) consisted of radial
velocities for only 43 clusters. Although these data were sufficient to provide a 
provisional mass estimate for M87, and to hint at the presence of 
rotation in the GCS, it was shown by Merritt \& Tremblay (1993) 
that this sample provides only weak constraints on the distribution of dark matter
in the M87 halo. Moreover, Merritt \& Tremblay (1993) demonstrated that a measurement 
of the velocity anisotropy of GCS would require radial velocities 
for a minimum sample of several hundred clusters. 

In a major observational effort, Cohen \& Ryzhov (1997) used the Keck telescope
to measure radial velocities for a total of 205 GCs surrounding 
M87. Using these data, Cohen \& Ryzhov (1997) found clear evidence for a rising 
velocity dispersion profile, and derived a total mass of 
$M = (3.8\pm1.2)\times10^{12} M_{\odot}$ within a volume of radius of 
$r \simeq 33$ kpc centered on M87. An independent analysis of these velocities 
by Kissler-Patig \& Gebhardt (1998) suggested that the M87
GCS is rotating rapidly at large radii: $i.e.$, $\Omega R \simeq 300$
km s$^{-1}$ beyond a projected radius of $R = 3R_e$ (20 kpc).\footnotemark
\footnotetext{According to de Vaucouleurs \& Nieto (1978), $R_e = 96^{{\prime}{\prime}}$
for M87. Throughout this paper, we adopt an M87 distance of $D = 15$ Mpc 
(Pierce et al. 1994; Ferrarese et al. 1996), so that 1$^{\prime}$ = 4.364 kpc.} 
This rapid rotation was subsequently confirmed by Cohen (2000) who augmented her
original radial velocity sample to include 17 additional clusters having 
$R \ge 3R_e$.

While the amount of angular momentum implied by these results is large, 
there is emerging evidence from radial velocity surveys of planetary nebulae 
that giant elliptical galaxies may, as a class, rotate rapidly at large radii 
(e.g., Hui et al. 1995; Arnaboldi et al. 1996; 1998). An enormous amount of
theoretical effort has been devoted to understanding the acquisition of 
angular momentum during galaxy formation: e.g., via tidal torques in
monolithic collapse scenarios (Peebles 1969), or through the conversion 
of orbital and spin angular momentum of the progenitor galaxies in major 
(Heyl, Hernquist \& Spergel 1996) and multiple (Weil \& Hernquist 1996) 
mergers. The angular momentum content of elliptical galaxies may thus 
constitute an important clue to their formation;
ultimately, the rotation properties of the chemically distinct GC
populations in these galaxies may provide a powerful means of 
discriminating between the various formation models described above.

In this paper, we present a dynamical analysis of the M87 GCS
using an improved and expanded sample of radial velocities. In addition,
new Washington CT$_1$ photometry is used to derive
metallicities for all program objects, and to investigate 
correlations between metallicity and dynamics by carrying out separate 
analyses for both the metal-rich and metal-poor GC populations.

\section{The Database}

The database used in the analysis presented below is a subset of that which
has been described in detail in a companion paper (Hanes et al. 2001; hereafter 
Paper I), so only a brief summary is given here. The reader is referred to 
Paper I for a complete description of the sample properties, observational 
material and data reductions.

\subsection{Photometry and Radial Velocities}

One of our goals is to investigate the relationship between GC dynamics 
and metallicity, so we require homogeneous metallicity estimates
for our sample of clusters. Such information is most easily obtained from
broadband photometry. Unfortunately, although there have been numerous 
photometric studies of the M87 GCs, the large areal coverage of our radial 
velocity survey means that no single existing survey includes all of our 
program objects. For instance, the most extensive publsihed photometric 
survey of the M87 GCS remains that of Strom et al. (1981),
who presented photographic $UBR$ photometry for 1728 candidate GCs
surrounding M87. Yet our sample includes many objects not included
in the Strom et al. (1981) catalog: e.g., objects that are at large radii, beyond the 
limits of their survey, or close to the galaxy center where crowding and
background contamination is severe. Therefore, as described in Paper I, we
have obtained deep, wide-field CCD images centered on M87 using the CTIO 
and KPNO 4m telescope. Imaging was carried out in Washington C and T$_1$ 
filters, since the long color baseline of the C$-$T$_1$ index provides 
exceptional metallicity sensitivity (see, e.g., Geisler, Lee \& Kim 1996). 
The photometric observations, calibrations, and reductions have been 
described briefly in Paper I, while a more comprehensive description of the 
photometric properties of the M87 GCS based on these data will be presented 
in a forthcoming paper (Geisler, Lee \& Kim 2001).

The sample of radial velocities we begin with here consists of data for
334 candidate and previously confirmed GCs. This number includes 145 measurements from
the Canada-France-Hawaii telescope (CFHT), 87 of which are for targets not
observed in any of the previous surveys of Huchra \& Brodie (1987), Mould, Oke
\& Nemec (1987), Mould et al.~(1990), Cohen \& Ryzhov (1997), or Cohen (2000).
The total is down slightly from the full database of 352 objects in Paper I,
as we have discarded out of hand 12 targets for which we were unable
to obtain unambiguous Washington photometry (ID nos.~881, 5053, 5055, 5058,
5064, 5065, 5066, 5067, 5071, 9002, 9051, and 9052 from Paper I); five which were already
identified by Cohen \& Ryzhov (1997; CR97) as Galactic M stars (ID nos.~38, 87, 782,
1551, 8001); and one object which turns out to be visibly non-stellar
(ID no.~207).

All measurements have been transformed onto common astrometric and radial
velocity scales. We adopt the radial velocity and uncertainty from 
our own analysis for all targets observed at the CFHT and transform all other data 
to this system as described in Paper I. Briefly, cluster
candidates that were not observed at the
CFHT, but were observed at Keck by Cohen \& Ryzhov (1997), are transformed onto our
MOS velocity scale according to:
$$ v_p({\rm MOS})=0.96\,v_p({\rm CR97}) + 91 $$
and assigned an uncertainty of $\pm 100$ km s$^{-1}$, as recommended by these authors.\footnotemark
\footnotetext{Here and throughout, quantities referred to the line of sight are denoted
by a subscript $p$.}
Objects measured neither by us nor by Cohen \& Ryzhov
(1997), but by Cohen (2000), are transformed as:
$$ v_p({\rm MOS})=0.98\,v_p({\rm C2000}) + 120 $$
and given an uncertainty of $\pm 50$ km s$^{-1}$. Finally, to targets with
velocities recorded {\it only} by Mould et al.~(1987) or Mould et al.~(1990),
we apply the correction:
$$ v_p({\rm MOS})=0.76\,v_p({\rm Mould}) + 280 $$
and adopt an uncertainty of $\pm 200$ km s$^{-1}$.

\begin{figure*}[t]
\centering \leavevmode
\epsfysize=4.0truein
\epsfbox{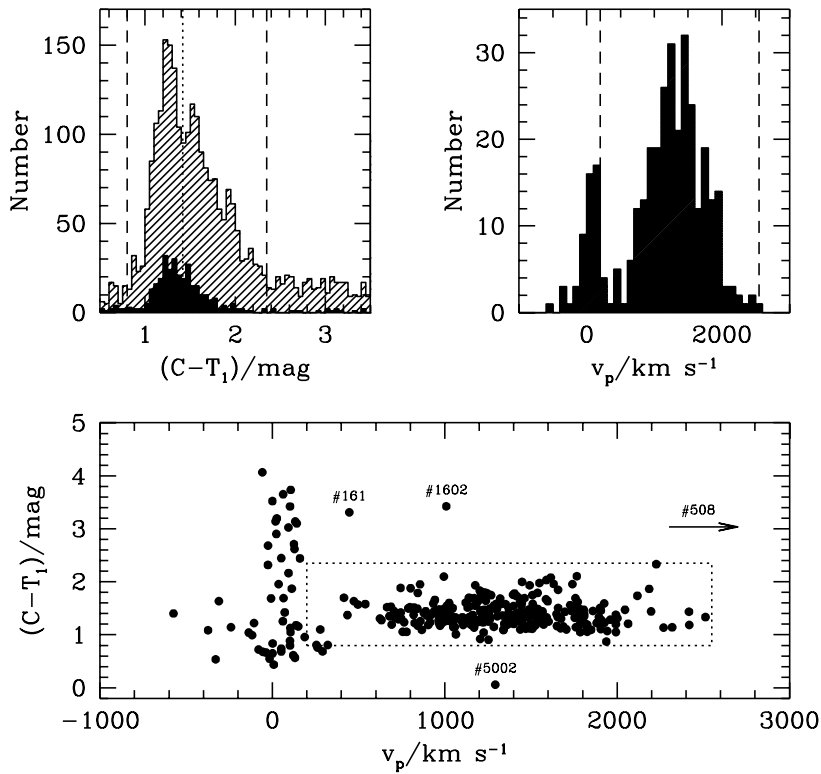}
\caption{{\it Upper Left Panel}: Distribution of (C$-$T$_1$) colors
for 334 globular cluster candidates with measured radial velocities (solid histogram).
For comparison, the color distribution of all 2744 globular cluster candidates
from the photometric study of Geisler, Lee \& Kim (2001) is shown as the dashed
histogram. The dashed vertical lines indicate our adopted selection criteria on color,
0.80 $\le$ (C$-$T$_1$) $\le$ 2.35 mag, while the dotted vertical line shows the color,
(C$-$T$_1$) = 1.42, used to isolate the blue and red cluster samples.
{\it Upper Right}: Radial velocity
histogram for the 334 candidate globular clusters with accurate (C$-$T$_1$)
colors. The vertical lines show our adopted selection criteria on velocity:
$200\le v \le 2550$ km s$^{-1}$. {\it Lower Panel}: Color vs.~radial velocity
for our initial sample of 334 objects. The dotted region shows the joint
selection criteria on color and radial velocity. This box defines a final
sample of 278 {\it bona fide} globular clusters.
\label{fig1}}
\end{figure*}

\subsection{Sample Selection}

Before proceeding with the dynamical analysis, it is important to first identify a sample of 
{\it bona fide} M87 GCs. Culling Galactic field stars from the sample on
the basis of radial velocity is not trivial since the relatively low systemic 
velocity of M87 ($1277\pm2$ km s$^{-1}$; van der Marel 1994) and the 
high velocity dispersion of the GCs (i.e., ${\langle}{{\sigma}_p}{\rangle} \sim 400$
km s$^{-1}$; see \S3) make the correct choice of the low velocity cutoff
problematical. We adopted a selection criterion of:
\begin{equation}
200 \le v_p \le 2550~{\rm km~s}^{-1}.
\label{eq1}
\end{equation}
This range is roughly symmetric about the mean velocity, $\langle v_p \rangle = 1350$
km s$^{-1}$, of the M87 GCS (see \S 3) and is expected to exclude the vast majority 
of foreground stars and background galaxies.
We also imposed a selection on C$-$T$_1$ color, discarding all objects which
did not have colors within the range:
\begin{equation}
0.8 \le {\rm C-T}_1 \le 2.35~{\rm mag}.
\label{eq2}
\end{equation}
The Galactic foreground extinction in the direction of M87, according to the 
DIRBE maps of Schlegel, Finkbeiner \& Davis (1998), is E(B-V) = 0.022 mag, which 
corresponds to E(C$-$T$_1$) = 0.045 mag (Secker et al. 1995). The metallicity 
calibration of Geisler \& Forte (1990) suggests that this 
color selection will confine our sample to clusters having metallicities
in range $-2.6 \le {\rm [Fe/H]} \le  1.0~{\rm dex}$, an interval which should encompass
virtually all GCs in M87 (e.g., Cohen, Blakeslee \& Ryzhov 1998). 
These selection criteria are illustrated in the upper panels of Figure \ref{fig1}.
The filled histograms in these two panels show the color and velocity distributions
for all 334 candidate GCs with measured radial velocities and (C$-$T$_1$) colors.
The dashed vertical lines in each panel show the adopted upper and lower 
selection limits. The dotted vertical line in the upper left panel shows our
adopted dividing point between the metal-rich and metal-poor cluster populations:
C$-$T$_1$ = 1.42. This color, which corresponds to a metallicity of
${\rm [Fe/H]} = -1.16$ dex, was chosen based on the observed dip in the (C$-$T$_1$)
distribution of the 2744 globular cluster candidates from the photometric catalog 
of Geisler et al. (2001) which have T$_1$ $\le$ 22 (i.e., the dashed histogram 
in this panel).

\begin{figure*}[t]
\centering \leavevmode
\epsfysize=4.0truein
\epsfbox{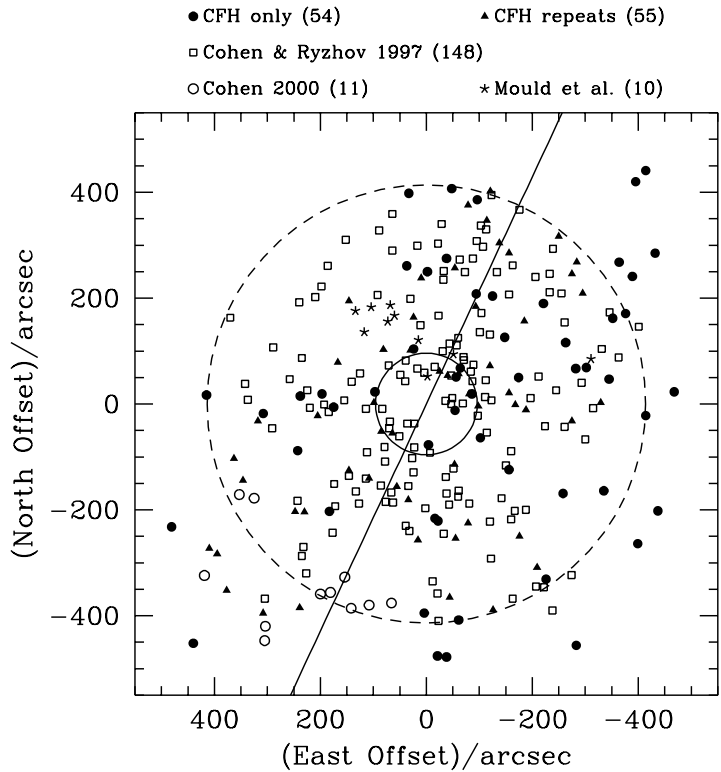}
\caption{Spatial distribution of the 278 confirmed globular
clusters around M87 (i.e., all objects falling within the boxed region
of the bottom panel of Figure~\ref{fig1}). The center of the galaxy is at
$(0,0)$, the inner circle marks the effective radius of the stellar mass
of M87 ($R_e = 96\arcsec\simeq 7$ kpc), and the outer circle has a
radius of $414\arcsec \simeq30\,{\rm kpc}\simeq 4.3 R_e$. The diagonal
line shows the photometric major axis of the galaxy.
\label{fig2}}
\end{figure*}

The lower panel of Figure~\ref{fig1} is a color-velocity diagram for these
same objects.
The dotted region shows the joint selection criteria on color and velocity,
which together identify {\it a total of 278 {\it bona fide} GCs}.
Of these, 161 are classified as ``blue'' (C$-$T$_1$ $<$ 1.42), or metal-poor, and
117 are considered ``red'' (C$-$T$_1$ $>$ 1.42), or metal-rich, for purposes of
the analysis and modelling in \S\S3 and 4 below.
Note that three objects (ID nos.~161, 508, and 1602) have radial velocities
which are  appropriate for M87 GCs, and yet have C$-$T$_1$ colors
which  are far too red for true clusters (i.e., the inferred metallicities
are typically several hundred times above solar). A fourth candidate (ID
no.~5002) has a radial velocity near the mean of the GC sample,
but an anomalously {\it blue} color. Although we have no reason
to suspect either the colors or the velocities of these objects,
their nature is unclear and additional observations are clearly
warranted, first and foremost to confirm our results (particularly for Strom
508, an unresolved object which has a high radial velocity of 
$v_p = 5817\pm120$ km s$^{-1}$ but a very red color of C$-$T$_1$ $\simeq$ 3.0). 
In any event, all four are omitted from the following analysis.

Our innermost GC is projected to a distance of roughly
$36\arcsec$ from the center of M87 (or about 2.6 kpc for our assumed
distance of 15 Mpc to the galaxy), while the outermost member of our sample
lies at $R\simeq 631\arcsec\simeq46$ kpc. Figure~\ref{fig2} shows the
distribution on the sky of all 278 {\it bona fide} GCs
(i.e., those objects falling inside the dotted region in the lower panel of
Figure~\ref{fig1}). The solid points on this plot denote 109
clusters which were observed by us at the CFHT; of these, 54 are entirely new
measurements and the rest are repeat observations of GCs identified in
one of the earlier studies. The open symbols refer to data taken from these 
other surveys, as indicated in the legend. The straight line running
through the figure represents the photometric major axis of the M87 halo light
(oriented at 155$^\circ$ East of North; Carter \& Dixon 1978), the inner
circle marks the effective radius of the galaxy light ($R_e = 96\arcsec
\simeq7$ kpc [de Vaucouleurs \& Nieto 1978]), and the larger, dashed
circle has a radius of
$414\arcsec \simeq30\,{\rm kpc}\simeq 4.3 R_e$. Outside this circle,
the azimuthal coverage of the composite radial-velocity database is far from
complete, a situation which can significantly compromise any inferences on the
kinematics of the GCS at such large galactocentric
distances (see below). 

\section{Kinematics of the Cluster System}

In previous analyses of the kinematics of the M87 GCS, investigators have
usually proceeded by fitting sine curves to their line-of-sight velocities, $v_p$,
as a function of projected azimuth, $\Theta$, in order to test for the presence of
rotation and to measure its amplitude (see Cohen \& Ryzhov 1997; Kissler-Patig
\& Gebhardt 1998; Cohen 2000) We shall do the same here, but we
pause first to clarify precisely what this assumes about the intrinsic
velocity field of the GCS.

\subsection{Mathematical Context}

Consider a {\it spherical} GCS rotating about some axis, with an angular
velocity $\omega$ that may be any axisymmetric function of position inside the
galaxy.\footnotemark
\footnotetext{While it is clearly
imperfect (see, e.g., McLaughlin, Harris, \& Hanes 1994), the assumption of
spherical symmetry in the M87 GCS is not grossly in error, and we invoke it as
a simplifying approximation throughout this paper.}
Let $r$, $\theta$, and $\phi$ denote the usual spherical coordinate
system within the GCS, such that $\theta=0$ naturally identifies its rotation
axis and its volume density $n$ is a function only of the galactocentric
distance $r$. Also define a projected, circular coordinate system on the plane
of the sky by the radial coordinate $R$ and the azimuth $\Theta$. The projected
(surface) density of the GCS, $N$, is, by the spherical hypothesis, a function
of $R$ alone:
\begin{equation}
N(R) = 2\int_{R}^{\infty} n(r) {{r\,dr}\over{\sqrt{r^2-R^2}}}\ .
\label{eq3}
\end{equation}
However, its {\it projected} angular velocity, $\Omega$, which is
just the average of $\omega$ along the line of sight through the GCS, may be a
function of $R$ and $\Theta$ both. Assuming that the cluster system is viewed
so that its rotation axis lies exactly in the plane of the sky and
coincides with $\Theta=0$, we have 
\begin{equation}
\Omega(R,\Theta) = {2\over{N(R)}}\,\int_{R}^{\infty} n(r)\omega(r,\theta)\,
{{r\,dr}\over{\sqrt{r^2-R^2}}}\ ,
\label{eq4}
\end{equation}
where $\cos\theta=(R\cos\Theta)/r$ by simple geometry (see also Fillmore 1986).
In addition to this,
the angular velocities $\omega$ and $\Omega$ are related to the linear
velocities $\langle v_\phi\rangle$ --- the intrinsic streaming, or rotation,
velocity of the GCS\footnotemark
\footnotetext{The $\phi$ direction wraps around the rotation axis by
definition.} --- and $\langle v_p \rangle$ --- the averaged component of velocity
along the line of sight --- in the usual way: $\omega=\langle v_\phi \rangle /
(r\sin\theta)$ and $\Omega=\langle v_p \rangle / R\sin\Theta$. Thus,
\begin{equation}
\langle v_p \rangle = 2\,{{R\sin\Theta}\over{N(R)}}\,
\int_{R}^{\infty} n(r) {{\langle v_\phi \rangle}\over{r\sin\theta}}\,
{{r\,dr}\over{\sqrt{r^2-R^2}}}\ ,
\label{eq5}
\end{equation}
where $\langle v_\phi\rangle$ may still, in general, be any function of
$r$ and $\theta$.

Equation (\ref{eq4}) makes it clear that the apparent angular velocity
will be a function of $R$ alone, $\Omega=\Omega(R)$, if $\omega$ is a
function only of $r$, i.e., if the intrinsic angular velocity of the GCS is
constant on spherical surfaces. This is just what is required, as equation
(\ref{eq5}) shows explicitly, to have $\langle v_p\rangle \propto \sin\Theta$,
and thus to justify fitting a sinusoid to our radial velocity data as a
function of projected azimuth.

It is worth emphasizing this point, even though it is not new (see, e.g.,
Fillmore 1986): the projected {\it angular} velocity of a spherical GCS is
constant on circles in the plane of the sky --- so that its linear velocity
$\langle v_p\rangle$ varies sinusoidally with projected azimuth $\Theta$ --- if
the intrinsic {\it angular} velocity is constant on spheres, such that 
$\langle v_\phi\rangle\propto \sin\theta$. It happens that the
variation of $\langle v_p\rangle$ with  $\Theta$ in our sample of M87 GCs
is adequately described, at all galactocentric radii, by a simple sine
curve (see below), and we interpret this as an indication that the angular
velocity field of the GCS is stratified approximately on spherical surfaces:
$\langle v_\phi\rangle/(r\sin\theta)=\omega(r)$. 
This interpretation is
somewhat different from that adopted by Cohen \& Ryzhov (1997), and apparently
by Kissler-Patig \& Gebhardt (1998), who fit sines to their $v_p$ vs.~$\Theta$
data and then posit that the intrinsic $\langle v_\phi\rangle$ field of the
GCS is stratified on {\it cylinders}. 
The only way that both of these assumptions can be simultaneously true ---
that is, $\langle v_p \rangle \propto \sin\Theta$ {\it and} $\langle
v_{\phi} \rangle$ depending on $r$ and $\theta$ only through the combination
$(r \sin \theta)$ --- is in the special case of solid-body rotation,
$\omega = \Omega = constant$.
This would, of course, require that $\Omega R \propto R$. 
Unfortunately, the data neither
rule out this possibility nor present a compelling case for it 
(see, e.g., Figure~\ref{fig8} below). We shall return to the issue of possible
solid-body rotation below, and in \S 5.7 for the specific case of the metal-poor 
GCS (see Figure~\ref{fig19}).

In summary, we assume that: (1) the GCS of M87 can be approximated as spherical
[$n=n(r)$ and $N=N(R)$]; (2) its angular velocity field is constant on
spheres [$\omega=\omega(r)$ and $\Omega=\Omega(R)$]; and (3) its rotation axis 
lies exactly in the plane of the sky and coincides with $\Theta=0$. 
We are then able to investigate the rotation of the GCS by
fitting our observed radial velocities with the function
\begin{equation}
v_p({\Theta}) = v_{\rm sys} + (\Omega R) \sin(\Theta - \Theta_0)\ ,
\label{eq6}
\end{equation}
where $\Theta_0$ (measured, as $\Theta$ is, in degrees East of North),
locates the GCS rotation axis, and where the rotation amplitude
$\Omega R$ can in principle be any function of projected galactocentric
radius $R$.

\subsection{Global Kinematic Properties}

\begin{figure*}[b]
\centering \leavevmode
\epsfysize=4.0truein
\epsfbox{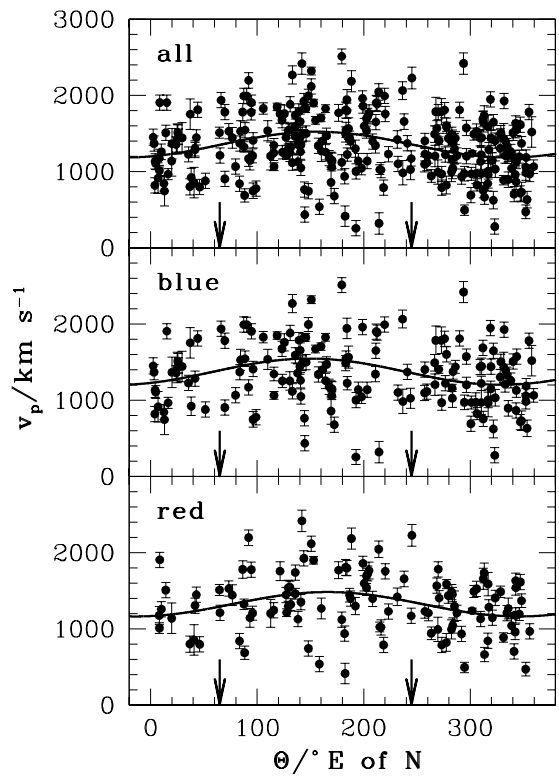}
\caption{Radial velocity vs.~azimuth for the full sample of 278
confirmed globular clusters, and separately for the 161 blue clusters with
0.80 $\le$ (C$-$T$_1$) $\le$ 1.42 and the 117 red globular clusters with 
1.42 $\le $ (C$-$T$_1$) $\le$ 2.35. The best-fit sine curve, 
$v_p = v_{\rm sys} + (\Omega R)\sin(\Theta - \Theta_0)$, is overlaid in each 
case; see Table \ref{tab1} for a listing of
the parameters. The orientation of the photometric minor axis of M87 is
indicated by the vertical arrows in each panel at $\Theta = 65^{\circ}$ and
$245^{\circ}$ East of North.
\label{fig3}}
\end{figure*}

Figure~\ref{fig3} shows the dependence of radial velocity on azimuth for
our full sample of 278 GCs in M87, for our subsample of
161 blue clusters (defined by 0.80 $\le$ C$-$T$_1$ $\le$ 1.42), and for our
subsample of 117 red GCs (1.42 $\le$ C$-$T$_1$ $\le$ 2.35). Overlaid on the
data in each case is the best-fit sine. 
For the full sample, we find a rotation amplitude of 
$\Omega R = 169^{+42}_{-97}$ km s$^{-1}$, where the quoted errors
refer to 90\% confidence intervals. This rotation amplitude is somewhat 
larger than the value of $\Omega R \sim 100$ km s$^{-1}$ estimated by 
Cohen \& Ryzhov (1997), although still consistent given the rather large 
uncertainties. The best-fit position angle of the rotation axis for
the full sample is $\Theta_0 = 66^\circ\pm35^\circ$ E of N.
The photometric minor axis of the underlying galaxy light is similarly
oriented to ${\Theta}_{\rm phot} = 65^{\circ}$ (Carter \& Dixon 1978), a
correspondence that is consistent with our simplifying assumption that we
are viewing the GCS essentially edge-on. We separately find $\Omega R =
172^{+51}_{-108}$ km s$^{-1}$ and $\Theta_0 = 59^\circ\pm52^\circ$
for the entire sample of metal-poor clusters, and  $\Omega R =
160^{+120}_{-99}$ km s$^{-1}$ and $\Theta_0 = 76^{\circ}\pm45^{\circ}$
for the metal-rich sample. The dispersion of velocities
about the best-fitting sine is $\sigma_{\rm p,r}=384^{+27}_{-32}$ km s$^{-1}$
for the full sample, $\sigma_{\rm p,r}=397^{+37}_{-46}$ km s$^{-1}$ for the
metal-poor clusters, and $\sigma_{\rm p,r}=364^{+49}_{-52}$ for the metal-rich 
clusters.

We conclude, therefore, that rotation is present in the M87 GCS and is
significant at a high confidence level, for the blue and red subsystems
alike. Moreover, the metal-poor and metal-rich clusters in our sample
have: (1) essentially identical rotation amplitudes; (2) rotation axes that 
coincide with one another and with the photometric minor axis of the galaxy; and
(3) similar fractions of their total kinetic energy stored in rotation
[$(\Omega R/\sigma_{\rm p,r})^2\simeq0.2\pm0.1$ in each case].
Thus, within the observational uncertainties, there is no obvious difference
between the {\it global} kinematics of the two subsystems. We caution, however, 
that this is just a thumbnail sketch of the kinematics, averaged over
the full range of galactocentric radii 
spanned by the GCs in our radial-velocity sample. As was stressed above,
the quantity $\Omega R$ can, in principle, be a function of projected
radius --- as might be the rotation axis $\Theta_0$ and the dispersion of the
line-of-sight velocities. While Figure~\ref{fig3} is useful as a rough
summary of the situation, it is necessarily limited.

\begin{figure*}[t]
\centering \leavevmode
\epsfysize=4.0truein
\epsfbox{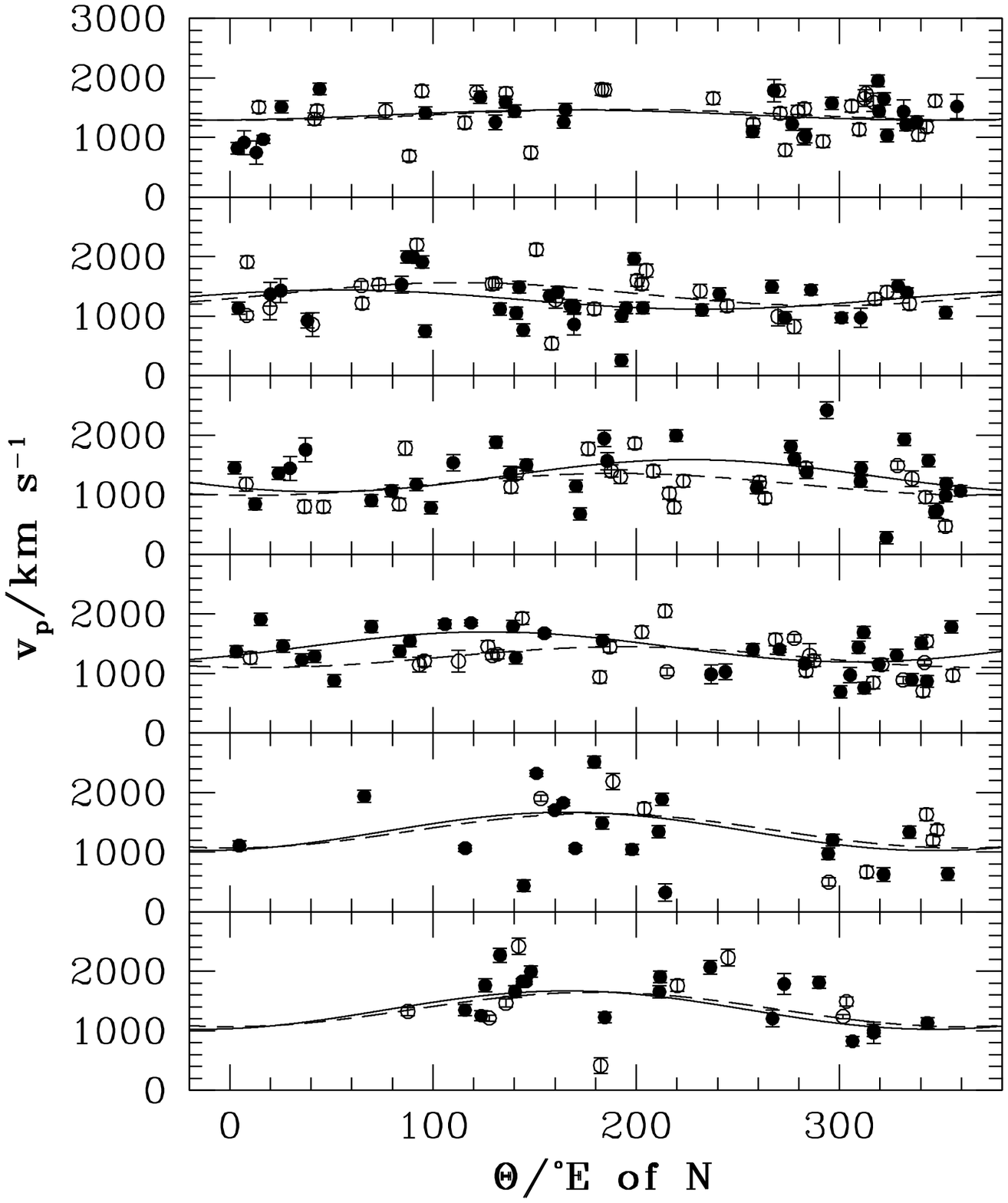}
\caption{Line-of-sight velocity vs.~azimuth for M87 globular clusters 
in six radial bins. From top to bottom, the panels show clusters in the range: 
(1)  $35\arcsec \le R < 125\arcsec$; 
(2) $125\arcsec \le R < 207\arcsec$;
(3) $207\arcsec \le R < 297\arcsec$;
(4) $297\arcsec \le R < 380\arcsec$;
(5) $380\arcsec \le R < 414\arcsec$;
(6) $414\arcsec \le R \le 635\arcsec$.
Filled circles are metal-poor (blue) globular clusters, and solid lines are 
their best-fit sine curves with parameters given in Table \ref{tab1}. Open 
circles denote the metal-rich (red) globular clusters, and broken lines trace 
the best-fit sine curves (with parameters also given in Table \ref{tab1}).
Note that the identical sine curves shown in the bottom two panels have been 
derived from the combined data: i.e., using all clusters in the range 
$380\arcsec \le R \le 635\arcsec$.
\label{fig4}}
\end{figure*}

Thus, we sort our GCs into a number of distinct circular annuli and fit
sine curves separately to $v_p$ vs.~$\Theta$ in each annulus to investigate
possible radial variations in the kinematics of the total, blue, and red
cluster systems. Table \ref{tab1} lists the results of this exercise and
includes full details of the global kinematics discussed just above.
The first column of the Table defines the annulus in question (with $1\arcsec=
0.073$ kpc for a distance of 15 Mpc to M87), the second and third columns gives
the median radius (in arcseconds and kpc) of the clusters in that radial bin, 
and the fourth records
the number of objects in it. The fifth and six columns then give the mean
line-of-sight velocity of the clusters (ignoring any rotation that might be
present, and estimated using the robust, biweight measure of ``location''
discussed by Beers, Flynn, \& Gebhardt 1990) and the dispersion of their
velocities about this mean value (i.e., the biweight ``scale'' of Beers et al. 1990).
The uncertainties on these values, and on all quantities in Table \ref{tab1},
define a 90\% confidence interval. They are estimated using a numerical
bootstrap procedure in which 1000 artificial datasets are constructed by
randomly choosing $N$ clusters from among the actual data in an annulus. The
mean and dispersion are computed for each of these 1000 trial datasets, the
results are sorted, and the values corresponding to the 5th and 95th percentiles
are identified; uncertainties are then defined as the offsets between these
values and the actual $\langle v_p\rangle$ and $\sigma_p$ computed from the read data. 
Columns 7, 8, and 9 of Table \ref{tab1} give the rotation parameters
in each annulus, for each of our GC subsamples. These are
determined by an error-weighted, nonlinear, least-squares fit of the sinusoid
in equation (\ref{eq6}), and their 90\% confidence intervals follow from
fitting $v_p$ vs.~$\Theta$ for each of the 1000 artificial datasets
constructed in the bootstrap procedure just described. Finally, column 10 of
Table \ref{tab1} gives the (biweight) dispersion of the velocities about the
best-fit sine in each annulus for comparison with the previous analyses
of Cohen \& Ryzhov (1997) and Kissler-Patig \& Gebhardt (1998).

\begin{deluxetable}{crrccccccc}
\scriptsize
\tablewidth{0pc}                        
\tablecaption{Kinematics of the M87 Globular Cluster System\label{tab1}}
\tablehead{                        
\colhead{$R$} &                       
\colhead{$\langle R\rangle$} &                      
\colhead{$\langle R\rangle$} &                      
\colhead{$N$} &                       
\colhead{$\langle v_p\rangle$} &                       
\colhead{$\sigma_p$} &
\colhead{$v_{\rm sys}$} &
\colhead{$\Theta_0$} &
\colhead{$\Omega R$} &                       
\colhead{${\sigma}_{\rm p,r}$} \nl
\colhead{(arcsec)} &
\colhead{(arcsec)} &
\colhead{(kpc)} &
\colhead{} &     
\colhead{(km s$^{-1}$)} &     
\colhead{(km s$^{-1}$)} &     
\colhead{(km s$^{-1}$)} &
\colhead{($^\circ\,$E of N)} &
\colhead{(km s$^{-1}$)} &                       
\colhead{(km s$^{-1}$)} }                        
\startdata                        
\multicolumn{9}{c}{} \nl
\multicolumn{9}{c}{Full sample: 278 clusters with $0.80\le (C-T_1)\le 2.35$} \nl
\multicolumn{9}{c}{} \nl
~35--635 & 252 & 18.3 & 278 & 1335$^{+~43}_{-~41}$ & 401$^{+~27}_{-~30}$ &
   1354$^{+~42}_{-~58}$ & ~66$^{+36}_{-34}$ & 169$^{+~42}_{-~97}$ &
   384$^{+~27}_{-~32}$ \nl
\multicolumn{9}{c}{} \nl
~35--125 & ~91 &  6.6 & ~55 & 1392$^{+~74}_{-~86}$ & 318$^{+~37}_{-~60}$ &
   1379$^{+~94}_{-~68}$ & ~92$^{+62}_{-70}$ & ~90$^{+130}_{-143}$ &
   306$^{+~41}_{-~58}$ \nl
125--207 & 178 & 12.9 & ~57 & 1282$^{+~93}_{-~90}$ & 366$^{+~66}_{-~82}$ &
   1312$^{+~63}_{-104}$ & ~$-1^{+56}_{-46}$ & 153$^{+133}_{-144}$ &
   359$^{+~58}_{-~89}$ \nl
207--297 & 252 & 18.3 & ~55 & 1269$^{+~93}_{-104}$ & 409$^{+~67}_{-~79}$ &
   1248$^{+129}_{-~61}$ & 127$^{+43}_{-55}$ & 188$^{+138}_{-122}$ &
   388$^{+~60}_{-~82}$ \nl
297--380 & 332 & 24.1 & ~56 & 1311$^{+~81}_{-~85}$ & 332$^{+~44}_{-~53}$ &
   1375$^{+~66}_{-122}$ & ~49$^{+45}_{-57}$ & 199$^{+113}_{-118}$ &
   301$^{+~39}_{-~49}$ \nl
380--635 & 415 & 30.2 & ~55 & 1445$^{+130}_{-129}$ & 543$^{+~73}_{-~88}$ &
   1360$^{+154}_{-119}$ & ~82$^{+36}_{-69}$ & 303$^{+190}_{-150}$ &
   470$^{+~96}_{-~97}$ \nl
\multicolumn{9}{c}{} \nl
380--414 & 402 & 29.2 & ~27 & 1324$^{+204}_{-203}$ & 605$^{+~98}_{-136}$ &
   1319$^{+232}_{-196}$ & ~45$^{+69}_{-49}$ & 361$^{+436}_{-221}$ &
   514$^{+105}_{-173}$ \nl
414--635 & 481 & 35.0 & ~28 & 1547$^{+172}_{-161}$ & 460$^{+~93}_{-114}$ &
   1381$^{+214}_{-~96}$ & 105$^{+45}_{-55}$ & 411$^{+305}_{-594}$ &
   352$^{+174}_{-118}$ \nl
\multicolumn{9}{c}{} \nl
\multicolumn{9}{c}{} \nl
\multicolumn{9}{c}{Metal-poor sample: 161 clusters with $0.80\le (C-T_1)\le 1.42$} \nl
\multicolumn{9}{c}{} \nl
~35--635 & 260 & 18.9 & 161 & 1339$^{+~58}_{-~63}$ & 412$^{+~36}_{-~43}$ &
   1375$^{+~43}_{-~92}$ & ~59$^{+50}_{-53}$ & 172$^{+~51}_{-108}$ &
   397$^{+~37}_{-~46}$ \nl
\multicolumn{9}{c}{} \nl
~35--125 & ~98 &  7.1 & ~26 & 1355$^{+126}_{-111}$ & 300$^{+~58}_{-~88}$ &
   1375$^{+109}_{-~95}$ & ~77$^{+76}_{-75}$ & ~85$^{+260}_{-166}$ &
   288$^{+~47}_{-104}$ \nl
125--207 & 185 & 13.5 & ~33 & 1235$^{+112}_{-116}$ & 350$^{+~96}_{-108}$ &
   1277$^{+~91}_{-125}$ & $-37^{+58}_{-54}$ & 162$^{+210}_{-229}$ &
   342$^{+~77}_{-117}$ \nl
207--297 & 250 & 18.2 & ~33 & 1325$^{+134}_{-143}$ & 447$^{+~83}_{-107}$ &
   1318$^{+180}_{-~93}$ & 136$^{+45}_{-52}$ & 270$^{+181}_{-182}$ &
   410$^{+~83}_{-120}$ \nl
297--380 & 334 & 24.3 & ~31 & 1353$^{+119}_{-106}$ & 340$^{+~50}_{-~71}$ &
   1445$^{+~36}_{-164}$ & ~37$^{+31}_{-55}$ & 253$^{+126}_{-141}$ &
   295$^{+~43}_{-~86}$ \nl
380--635 & 414 & 30.1 & ~38 & 1433$^{+166}_{-172}$ & 531$^{+~81}_{-106}$ &
   1346$^{+156}_{-152}$ & ~76$^{+46}_{-64}$ & 322$^{+229}_{-175}$ &
   452$^{+~98}_{-128}$ \nl
\multicolumn{9}{c}{} \nl
380--414 & 404 & 29.4 & ~19 & 1291$^{+239}_{-259}$ & 600$^{+126}_{-235}$ &
   1280$^{+225}_{-282}$ & ~63$^{+75}_{-77}$ & 313$^{+476}_{-337}$ &
   555$^{+112}_{-240}$ \nl
414--635 & 480 & 34.9 & ~19 & 1559$^{+263}_{-183}$ & 412$^{+~71}_{-171}$ &
   1406$^{+199}_{-140}$ & ~94$^{+29}_{-45}$ & 443$^{+316}_{-252}$ &
   303$^{+~61}_{-153}$ \nl
\multicolumn{9}{c}{} \nl
\multicolumn{9}{c}{} \nl
\multicolumn{9}{c}{Metal-rich sample: 117 clusters with $1.42\le (C-T_1)\le 2.35$} \nl
\multicolumn{9}{c}{} \nl
~35--635 & 239 & 17.4 & 117 & 1331$^{+~63}_{-~66}$ & 385$^{+~49}_{-~51}$ &
   1324$^{+~93}_{-~70}$ & ~76$^{+44}_{-46}$ & 160$^{+120}_{-~99}$ &
   364$^{+~49}_{-~52}$ \nl
\multicolumn{9}{c}{} \nl
~35--125 & ~83 &  6.0 & ~29 & 1415$^{+137}_{-121}$ & 325$^{+~70}_{-~95}$ &
   1379$^{+153}_{-116}$ & 100$^{+71}_{-78}$ & ~93$^{+178}_{-256}$ &
   319$^{+~61}_{-137}$ \nl
125--207 & 172 & 12.5 & ~24 & 1341$^{+156}_{-133}$ & 378$^{+~95}_{-138}$ &
   1368$^{+116}_{-155}$ & ~23$^{+55}_{-62}$ & 193$^{+205}_{-246}$ &
   345$^{+~88}_{-155}$ \nl
207--297 & 258 & 18.8 & ~22 & 1199$^{+131}_{-137}$ & 344$^{+~79}_{-114}$ &
   1174$^{+139}_{-116}$ & 100$^{+53}_{-63}$ & 183$^{+195}_{-115}$ &
   312$^{+~44}_{-138}$ \nl
297--380 & 325 & 23.6 & ~25 & 1257$^{+106}_{-119}$ & 304$^{+~83}_{-120}$ &
   1271$^{+116}_{-112}$ & 104$^{+43}_{-55}$ & 174$^{+306}_{-199}$ &
   268$^{+~55}_{-141}$ \nl
380--635 & 415 & 30.2 & ~17 & 1499$^{+207}_{-275}$ & 588$^{+118}_{-276}$ &
   1357$^{+324}_{-232}$ & ~90$^{+67}_{-73}$ & 289$^{+441}_{-595}$ &
   477$^{+187}_{-270}$ \nl
\multicolumn{9}{c}{} \nl
380--414 & 399 & 29.0 & ~~8 & 1433$^{+407}_{-387}$ & 583$^{+114}_{-442}$ &
   1350\tablenotemark{a} & ~18$^{+73}_{-50}$ & 667$^{+~487}_{-1000}$ &
   314$^{+112}_{-314}$ \nl
414--635 & 481 & 35.0 & ~~9 & 1512$^{+284}_{-287}$ & 568$^{+196}_{-568}$ &
   1350\tablenotemark{a} & 116$^{+66}_{-58}$ & 200$^{+~973}_{-1050}$ &
   432$^{+360}_{-432}$ \nl
\enddata                        
\tablenotetext{a}{Systemic velocity held fixed in fit of sine curve.}
\end{deluxetable}

Figure~\ref{fig4} directly compares some of the fits from Table \ref{tab1}
against the data. The filled circles in this figure plot $v_p$ against
$\Theta$ for our blue GCs, while the open circles correspond to the red
clusters. The top panel shows the objects with $35\arcsec \le R <
125\arcsec$; the second panel is for those with $125\arcsec \le R <
207\arcsec$; the next panel includes clusters at $207\arcsec \le R <
297\arcsec$; the fourth is for GCs with $297\arcsec \le R <
380\arcsec$; the fifth has $380\arcsec \le R < 414\arcsec$; and the sixth,
$414\arcsec \le R \le 635\arcsec$. In the top four panels, the solid curves
trace the corresponding sine fits for the blue clusters, as summarized in Table
\ref{tab1}, and the broken curves are the best-fit sinusoids for the red
clusters. In both of the bottom two panels, the blue and red fits shown are
those given in Table \ref{tab1} for the annuli {\it combined} (i.e., for
$380\arcsec\le R \le 635\arcsec$), as these are much more stable and
less uncertain. In all cases, the observed $v_p(\Theta)$ data appear to be
consistent with the simple sinusoidal form that was assumed in our fitting.

\begin{figure*}[b]
\centering \leavevmode
\epsfysize=4.0truein
\epsfbox{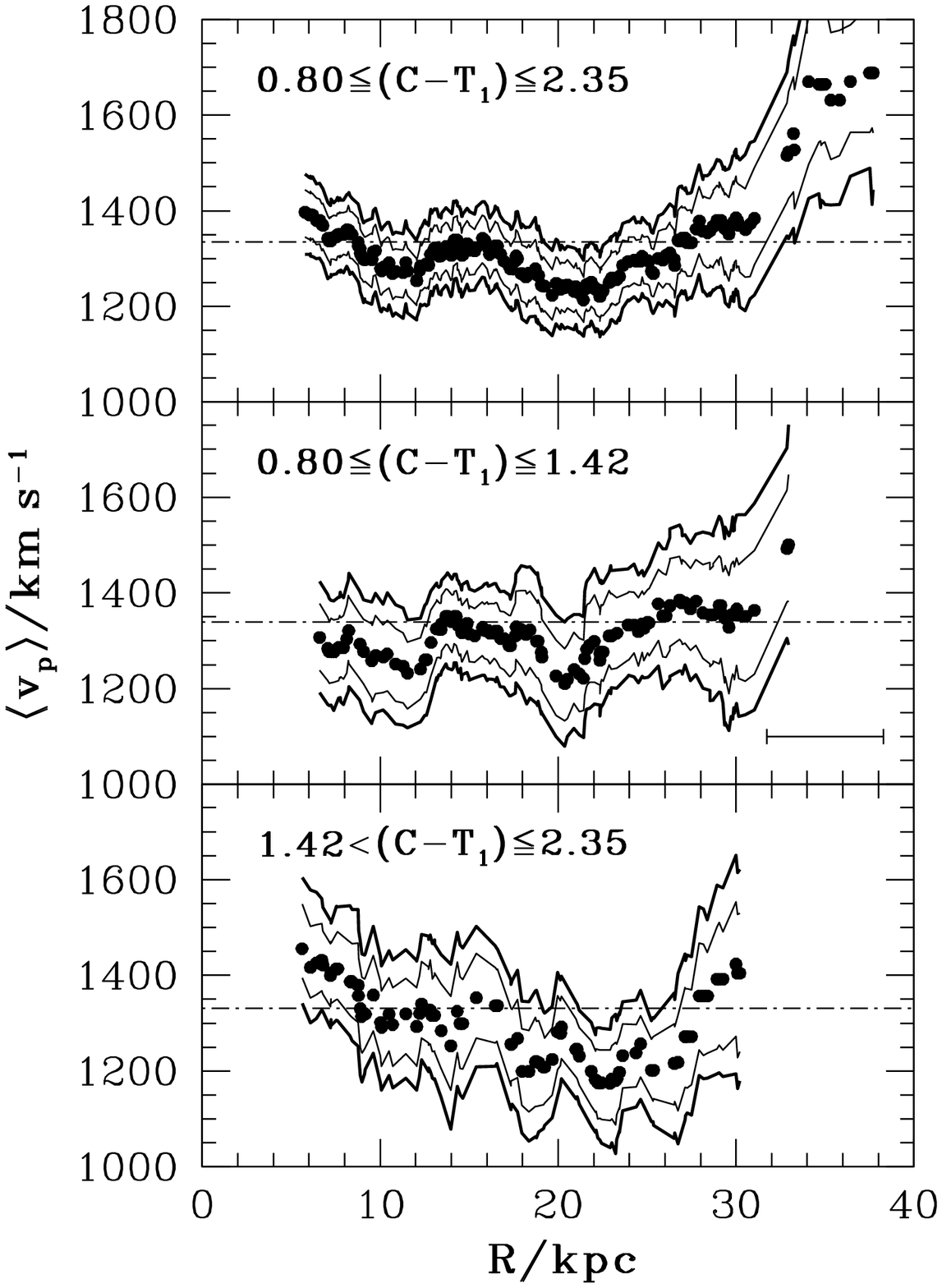}
\caption{{\it Upper Panel}: Biweight estimates for the mean
line-of-sight velocity of the M87 globular cluster system, as a 
function of distance from galaxy center (filled circles).  Thin and thick
solid lines show the 68\% and 90\% confidence limits on $\langle v \rangle$,
estimated using a bootstrap technique as described in the text. Points
represent the systemic velocity measured in radial bins of constant width
${\Delta}R \equiv 90\arcsec\simeq 6.5$ kpc (indicated by the error bar in 
the middle panel). The broken horizontal line indicates the mean velocity
recorded in Table \ref{tab1} for clusters at all radii $35\arcsec\le R\le
635\arcsec$. {\it Middle Panel}: Same as above, but for the sample of
metal-poor (blue) clusters. {\it Lower Panel}: As above, but for the sample of
metal-rich (red) clusters.
\label{fig5}}
\end{figure*}

Table \ref{tab1} and Figure~\ref{fig4} hint at some interesting radial trends
in the kinematics of the M87 GCS. First, it appears that the projected 
kinematics of the red and blue subsystems are, at least to first order, rather similar.
There are, however, suggestions of some differences which we
shall return to below and in \S 5.7 where we discuss the two-dimensional
velocity field of the GCS. Second,
although it is roughly constant over much of our range in projected radius,
the velocity dispersion of the clusters (either about the sample mean $\langle
v_p\rangle$ or about the best-fit sine curve) shows some evidence for an
increase at the largest $R$. This is just significant at the 90\% level in our
sample (and not even then for the metal-rich subsystem), but it is actually
predicted by the simple dynamical models we present in \S4. Third, there is
some evidence for a rise in the rotation amplitude $\Omega R$ at large radii, although
this is not formally significant at the 90\% level; to this degree of accuracy,
the red and blue cluster samples both are consistent with a constant
$\Omega R\sim 170$ km s$^{-1}$ at all radii $2.5\la R\la 35$ kpc (see
Figure~\ref{fig8} below). On the other hand, the angle of the rotation axis, 
$\Theta_0$, {\it does} change significantly as a function of radius, particularly
among the blue clusters. This interesting phenomenon is discussed in more
detail below.

\begin{figure*}[t]
\centering \leavevmode
\epsfysize=4.0truein
\epsfbox{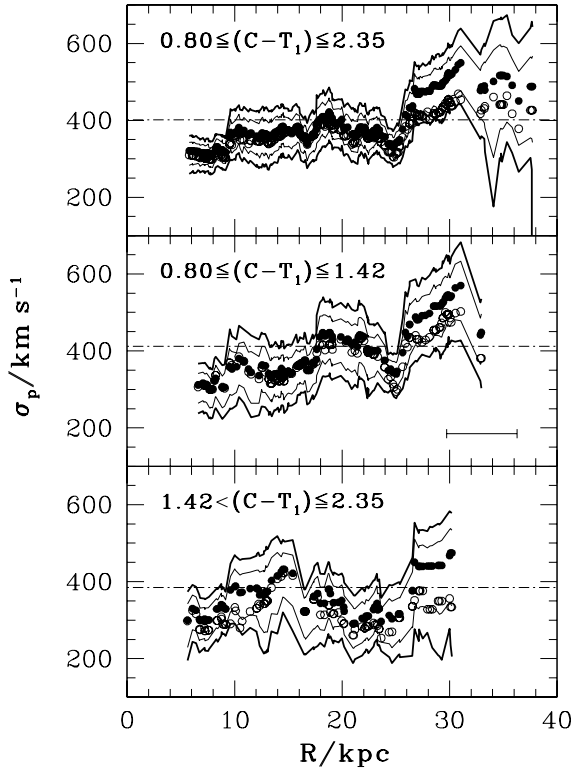}
\caption{{\it Upper Panel}: Biweight estimates for the
velocity dispersion of the M87 globular cluster system as a function of galactocentric 
distance.  Solid points denote the dispersion $\sigma_p$ about the sample mean in each
bin, taken from Figure~\ref{fig5}. Open circles represent the dispersion
$\sigma_{\rm p,r}$ about the best-fitting sine curve in each case (parameters
taken from Figures~\ref{fig7} and \ref{fig8}). Solid curves delimit the 68\% and
90\% confidence bands for $\sigma_p$, obtained from a numerical bootstrap.
Horizontal line is at the value determined for the full sample of clusters
with $35\arcsec\le R\le 635\arcsec$. {\it Middle Panel}: As above, but for the
sample of metal-poor (blue) globular clusters. {\it Lower Panel}: As above, for the
sample of metal-rich (red) clusters.
\label{fig6}}
\end{figure*}

\subsection{Kinematic Properties as a Function of Projected Radius}

\begin{figure*}[t]
\centering \leavevmode
\epsfysize=4.0truein
\epsfbox{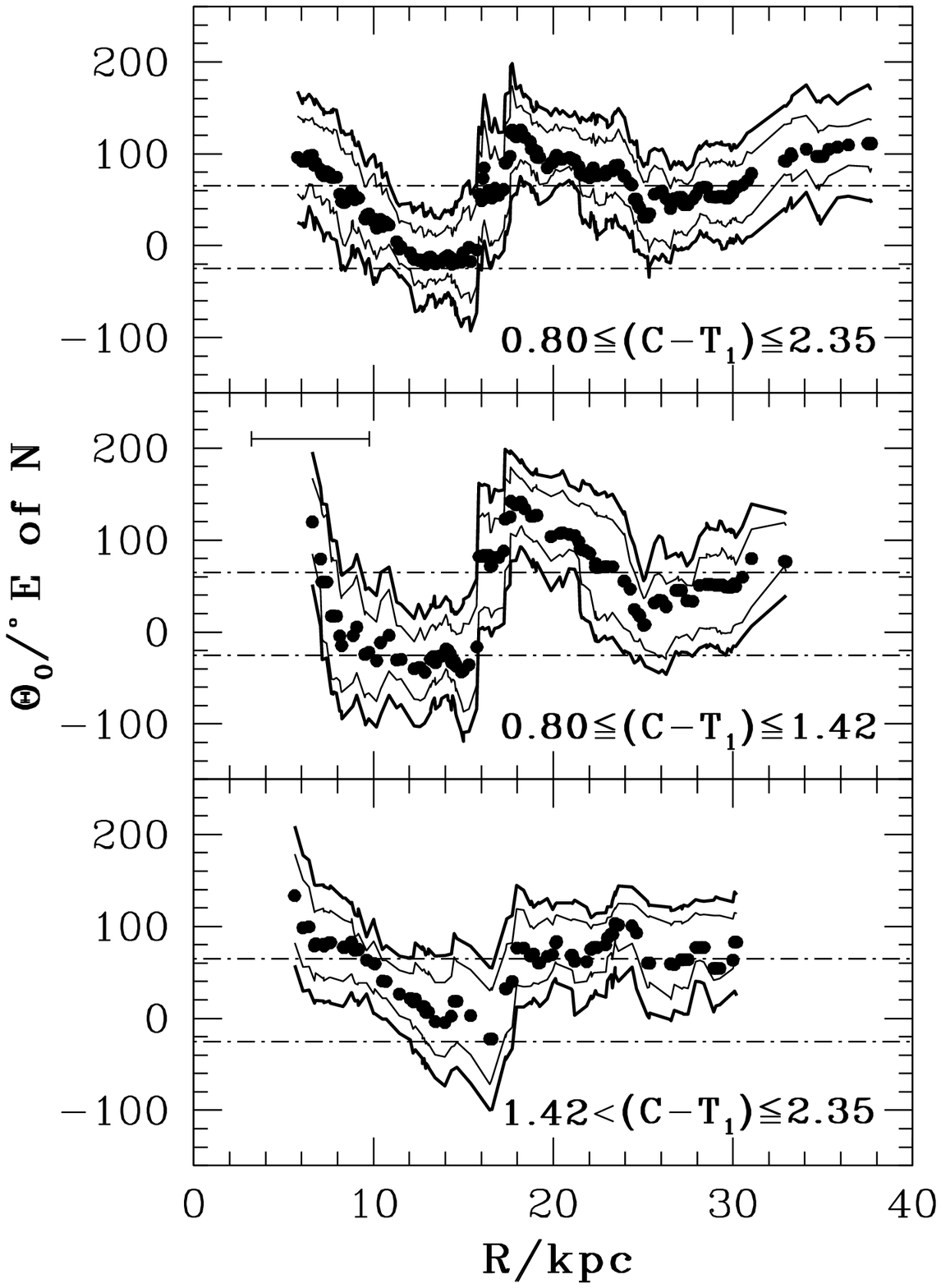}
\caption{{\it Upper Panel}: Projected azimuth of the rotation
axis of the full M87 globular cluster system, as a function of projected galactocentric radius
(points). Thin and thick solid lines define 68\% and 90\% confidence bands. In
this figure, and the following one, the systemic velocity was held fixed at 
$v_{\rm sys}\equiv 1350$ km s$^{-1}$ (see Table \ref{tab1}). The upper and
lower horizontal lines indicate position angles of the projected major ($\Theta_0=-25^\circ$) 
and minor ($\Theta_0=65^\circ$) axes of M87.
{\it Middle Panel}: For the metal-poor (blue) globular clusters. 
{\it Lower Panel}: For the metal-rich (red) globular clusters.
\label{fig7}}
\end{figure*}

\begin{figure*}[b]
\centering \leavevmode
\epsfysize=4.0truein
\epsfbox{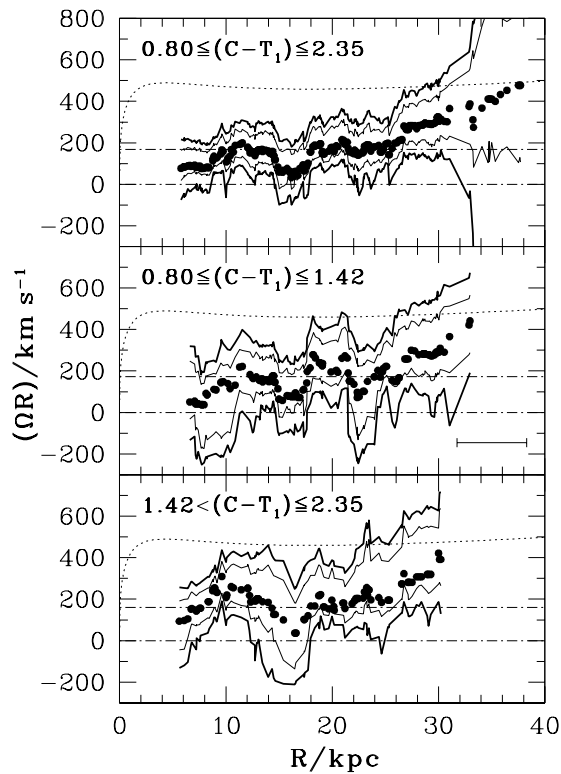}
\caption{{\it Upper Panel}: Amplitude of projected
rotation as a function of galactocentric radius in the M87 globular cluster system (filled
circles), shown with 68\% and 90\% confidence bands (thin and thick solid
lines). The position angle, $\Theta_0$, of the rotation axis was allowed to
vary with radius in the sinusoidal fits (see Figure~\ref{fig7}), but the
systemic velocity was held fixed at $v_{\rm sys}\equiv 1350$ km s$^{-1}$ (see
Table \ref{tab1}). Upper horizontal (dash-dot) line is placed at the amplitude
$\Omega R = 169$ km s$^{-1}$ found by fitting to the entire sample of globular clusters, over all radii
(see Table \ref{tab1}). Dotted curve shows the circular velocity as a function
of radius in the M87/Virgo gravitational potential, $v_c(r)=[GM(r)/r]^{1/2}$.
{\it Middle Panel}: As above, but for the sample of metal-poor (blue) clusters.
{\it Lower Panel}: As above, but for the sample of metal-rich (red) clusters.
\label{fig8}}
\end{figure*}

In order to better visualize these trends, and to address
any concerns that our particular choice of bins in Table \ref{tab1} might have
influenced our results, Figures~\ref{fig5}--\ref{fig8} show the line-of-sight
mean velocity and dispersion, and the best-fit rotation axis and amplitude, as
functions of projected radius for our full GC sample, and
separately for each of the blue and red subsets. To construct these profiles
we slide a bin of fixed radial width $\Delta R=90\arcsec\simeq6.5$ kpc through
the GCS, centering the bin at the position of each of our measured GCs
in turn and computing the kinematics (including bootstrap estimates of
confidence intervals) for all clusters lying within $\pm(\Delta R)/2$ of that
point. We begin at small projected radii when the number of clusters per bin
first exceeds 15 (so that the fits can be usefully constrained) and we stop at
large $R$ when the sample size per bin drops below this limit.

\begin{figure*}[t]
\centering \leavevmode
\epsfysize=4.0truein
\epsfbox{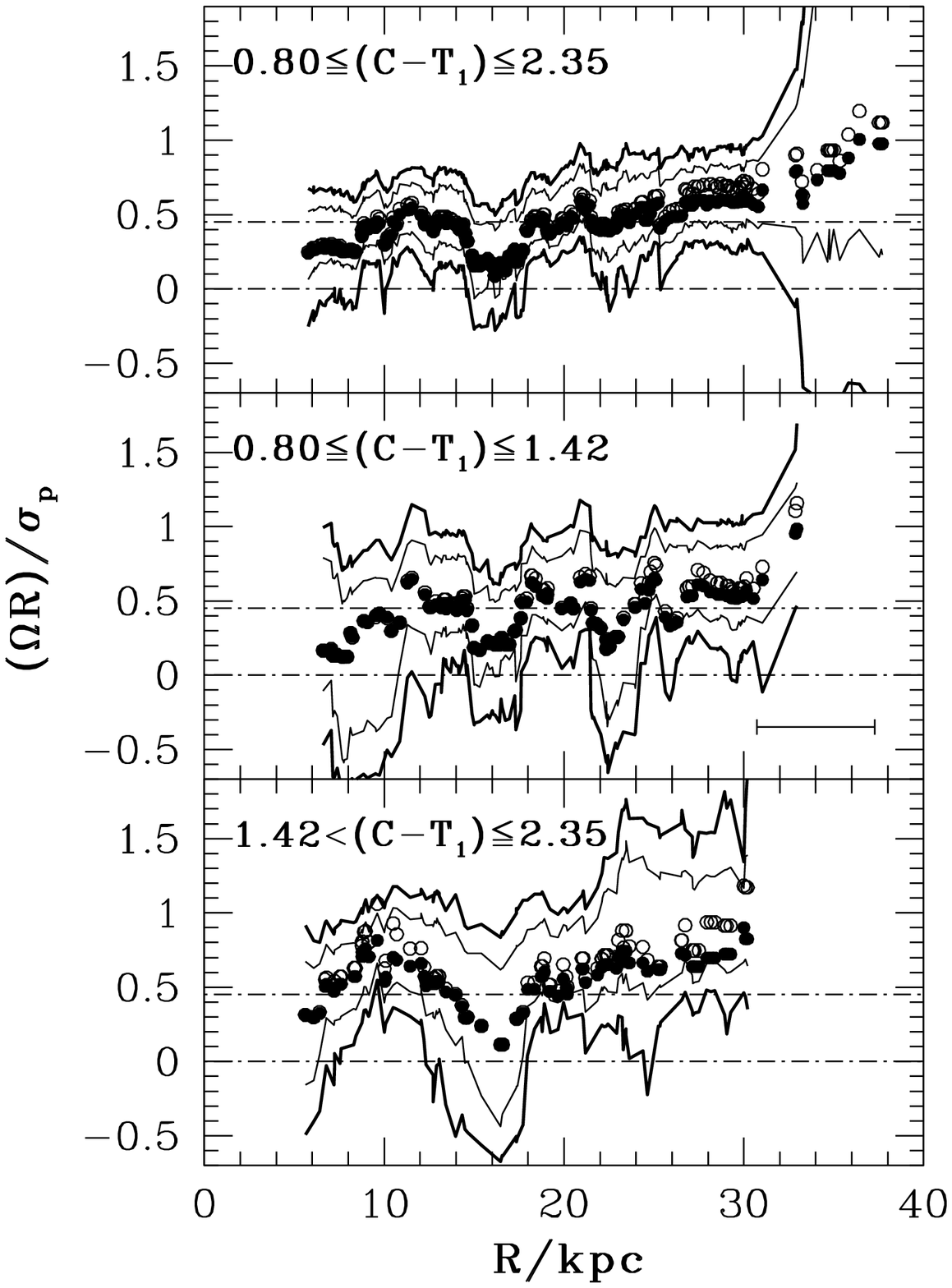}
\caption{{\it Upper Panel}: Ratio of the projected rotation
amplitude in Figure~\ref{fig8} to the line-of-sight velocity dispersion in
Figure~\ref{fig6}. Solid points compare $\Omega R$ to the biweight dispersion
$\sigma_p$ about the mean velocity $\langle v_p\rangle$ at each radius, and
the solid lines trace the 68\% and 90\% confidence limits on the ratio. Open
points come from comparing $\Omega R$ to the biweight dispersion
$\sigma_{\rm p,r}$ about the best-fitting sine curve at each radius. For
clarity, the confidence bands for this second ratio are not shown. The upper 
horizontal line marks the average of $\langle{({\Omega R})/{{\sigma}_p}}\rangle$ = 0.45 
suggested by a fit
to all cluster in our sample, regardless of radius (see Table \ref{tab1}).
{\it Middle Panel}: As above, but for the metal-poor (blue) globular clusters.
{\it Lower Panel}: As above, for the metal-rich (red) globular cluster system.
\label{fig9}}
\end{figure*}

Clearly, most of the points plotted in these figures are not statistically
independent---only points separated by at least the full $\Delta R=6.5$ kpc
(the size of the horizontal bar drawn in each middle panel) can be---and we
do not attempt any sort of formal fits to these profiles. Nevertheless, this 
procedure is useful since it provides a more detailed (albeit smoothed) view
of radial trends than can be had with a single binning such as that in Table
\ref{tab1}. It is also important to note that this method of
smoothing the data is different than that employed by Kissler-Patig \&
Gebhardt (1998) and Cohen (2000). These authors opt instead to fit the
kinematics in a series of annuli that always enclose the same number of
data points, and which therefore (because neither the intrinsic spatial
distribution of the GCS nor the observational sampling of it is uniform) have
a {\it variable} radial width (for an illustration of this point, see Figure 2 of 
Cohen 2000). Our approach avoids the nontrivial difficulty of interpreting data
that have been smoothed by different amounts from point to point, and it
yields a more accurate, though necessarily less precise, depiction of
radial variations in kinematics at the low-density, large-radius limits of
our GC sample.

The mean velocity in Figure~\ref{fig5}, which makes no allowance or
corrections for rotation, shows no significant variations with
radius until roughly $R\ga 30$ kpc. The sharp apparent rise in $\langle
v_p\rangle$ at this radius is due --- as Figure~\ref{fig2} and the bottom of 
Figure~\ref{fig4} both show, and as Romanowsky \& Kochanek (2000) also
suggested --- to sparse sampling of a population that shows significant
rotation: the few GCs in our radial-velocity
sample at these large radii have been drawn mostly from positions around
the photometric major axis of M87 (at 155$^\circ$ East of North), where the
rotation produces line-of-sight velocities furthest from the true systemic
velocity of the galaxy and the GCS. Fitting a sine curve to quantify the
rotation at every radius takes this effect into account, and a plot
of $v_{\rm sys}$ vs.~$R$ (cf.~eq.~[\ref{eq6}]) is essentially flat, and
consistent with a constant $v_{\rm sys}\equiv 1350$ km s$^{-1}$.

The velocity dispersion profiles in Figure~\ref{fig6} show both the biweight
rms scatter, $\sigma_p$, about the direct mean of the sample in each bin 
(solid points) and the biweight dispersion, $\sigma_{\rm p,r}$, about
the best-fitting sine curve (open circles). At relatively small projected
radii $R\sim10$--15 kpc, the
red clusters may be dynamically somewhat ``hotter'' than the blue
clusters in M87, and the situation may be reversed at larger radii.
As was mentioned above, however, in connection with Table \ref{tab1}, these
apparent differences are not statistically significant.
However, in agreement with
previous studies (e.g., Cohen \& Ryzhov 1997), we do see evidence for a rising
velocity dispersion towards larger radii --- a trend that is more clearly
established for our blue clusters than for the red ones.

Figures~\ref{fig7} and \ref{fig8} --- which show the rotation-axis position
angle and apparent rotation amplitude as functions of projected radius in the
M87 GCS --- present the most substantial new result of this empirical part of our
analysis and constitute a point of clear departure from earlier studies. These
profiles were obtained by fitting equation (\ref{eq6}) to the $v_p(\Theta)$
data in our sliding radial bin, with $v_{\rm sys}\equiv 1350$ km s$^{-1}$ held
fixed but with $\Theta_0$ and $\Omega R$ both allowed to vary. The
most striking point relates to the blue clusters, which appear (see
Figure~\ref{fig7}) to rotate about the projected {\it major} axis of M87
($\Theta_0=-25^\circ$ E of N) in the inner regions of the system,
$R\la 16$--18 kpc, but then switch to the the more conventional minor-axis
rotation ($\Theta_0=65^\circ$ E of N) that they show for $R\ga 20$ kpc. Given
the smoothing kernel that we have applied to the data (again,
$\Delta R=90\arcsec\simeq6.5$ kpc, as illustrated in the middle panel of
Figure~\ref{fig7}), this 90-degree flip in $\Theta_0$ seems to be almost
instantaneous. It is probably no accident that it appears to coincide
with the onset, at $R\simeq19$ kpc, of the cD envelope in the starlight
(de Vaucouleurs \& Nieto 1978; Carter \& Dixon 1978)
{\it and} the GCS of M87 (McLaughlin, Harris, \& Hanes 1993). 
Figures~\ref{fig7} and \ref{fig8} also reveal that the rotation axis of the 
red clusters seems to show a gradual
drift towards the photometric major axis of M87 for $R\la 18$ kpc before
it ``corrects'' to a constant $\Theta_0\simeq 65^\circ$ at larger radii. It is
not clear, however, that this apparent difference in the small-$R$ behavior of
the red and blue GCs is statistically significant --- even at the 68\%
confidence level. We return
to this issue in \S5 below, where the two-dimensional rotation fields of
the metal-poor and metal-rich GC samples are discussed within the context of 
the larger Virgo cluster environment.

Figure~\ref{fig8} shows the rotation amplitude corresponding to each fitted
value of $\Theta_0$ in Figure~\ref{fig7}. As was suggested above, it is
difficult to discriminate between a flat rotation curve and one which 
rises with increasing radius. For instance, there is some evidence for 
an outward rise in $\Omega R$ for $R\ga 25$ kpc, for the red and
blue GCs alike, but this effect sets in roughly where the azimuthal
coverage of our radial-velocity sample begins to decline (see Figures~\ref{fig2}
and \ref{fig4}), and it is significant at just the 68\% confidence level.
Also, note that the dashed curve drawn in all three panels of Figure~\ref{fig8}
shows the circular velocity of the M87/Virgo potential in which the GCs
are orbiting [$v_c(r)=\{GM(r)/r\}^{1/2}$; see eq.~(\ref{eq11}) below for
our adopted mass model]. The fact that $\Omega R$
appears to approach this limit at our largest sampled radii suggests that the
sparser spatial coverage there may be biasing our results to some
extent; it may also provide some evidence that the M87
GCS is being viewed nearly edge-on, as any inclination correction is likely
to be small if the apparent $\Omega R$ is already so near $v_c(r)$.

One other point worthy of note in these $\Omega R$ profiles is the apparent
dip to nearly 0 at radii $R\sim14$--18 kpc. This may simply be an artifact of
smoothing the data, rather than an indication of any real feature in the
intrinsic velocity field of the GCS. Since $\Theta_0$ appears to change very
suddenly by roughly 90$^\circ$ at some radius in this range, it is conceivable
that the 6.5-kpc wide annular bins centered on points in this part of the GCS
are simply combining two distinct samples of clusters, rotating about
essentially orthogonal axes, whose sinusoidal $v_p(\Theta)$ dependences
interfere destructively and average out to a lower amplitude of apparent
rotation.

Using the spectroscopic metallicities of Cohen, Blakeslee, \& Ryzhov (1998),
Kissler-Patig \& Gebhardt (1998) attempted to 
distinguish between metal-poor and metal-rich clusters in their study of 
the rotation properties of the M87 GCS.
They concluded that the blue clusters showed essentially {\it no}
rotation, $\Omega R\simeq0$, for small radii $R\la 15$ kpc. However, they 
fixed $\Theta_0$ at about 30$^\circ$ E of N in their analysis, and solved only for the
amplitude when fitting sines to their $v_p(\Theta)$ data. The difference
between their result and ours for the blue-GCS rotation amplitude stems
from the fact that we have found $\Theta_0$ at small $R$ to be far away (i.e.,
approaching 90$^\circ$) from the position angle adopted by
Kissler-Patig \& Gebhardt (1998).

The acquisition of still more radial velocities for M87 GCs
will be important in confirming our findings and in clarifying and quantifying
any dichotomies that might exist in the blue vs.~red kinematics.
Meanwhile, the surprisingly complicated rotation field that we have uncovered
demands the development of sophisticated dynamical models of the GCS. In
particular, although we proceed in the next Section with a standard,
one-dimensional Jeans-equation analysis of velocity moments in the M87 GCS, we
recognize that we cannot hope thereby to account rigorously for the dynamical
effects of rotation. In fact, in this analysis we shall neglect rotation
altogether, treating it as a perturbation to be addressed in future
studies.

Figure~\ref{fig9} roughly quantifies the effect this choice may have on our
results by comparing the rotation amplitude of the M87 GCS to its
line-of-sight velocity dispersion, both about the sample mean (filled circles)
and about the best-fit sine (open circles), at every projected radius. The
ratio of the two is essentially constant with projected radius, holding at a
level of about 0.45 for the GCS as a whole and for the metal-poor and
metal-rich subsamples individually. Thus, rotation enters as a correction at
the $(\Omega R/\sigma_p)^2\sim20\%$ level in the Jeans-equation analysis that
we shall pursue below and, thus, at this exploratory stage, is a relatively
small source of uncertainty.

\placefigure{fig9}

\section{Dynamical Models}

In the absence of rotation and the approximation of spherical symmetry, the
fundamental equation of our dynamical analysis is the spherical Jeans
equation (e.g., Binney \& Tremaine 1987):
\begin{equation}
{d\over{dr}}\, n_{\rm cl} \sigma_r^2 +
{{2\,\beta_{\rm cl}}\over{r}}\, n_{\rm cl} \sigma_r^2 =
- n_{\rm cl}\,{{GM_{\rm tot}(r)}\over{r^2}}\ ,
\label{eq7}
\end{equation}
where $n_{\rm cl}(r)$ is the volume density profile of the GCS,
$\sigma_r(r)$ is its velocity dispersion in the radial direction,
and $\beta_{\rm cl}(r)\equiv 1-\sigma_\theta^2/\sigma_r^2$ is a
measure of its velocity anisotropy.\footnotemark
\footnotetext{Considerations of symmetry imply that
the two tangential velocity dispersions are equal, i.e., $\sigma_\theta
= \sigma_\phi$.}

The tendency in previous studies along these lines has been to use the
projected velocity dispersions of the GCs to infer the
intrinsic $\sigma_r(r)$ and thus to constrain the total, ``background''
gravitating mass profile $M_{\rm tot}(r)$. To do so, however, requires some
{\it a priori} assumptions on the anisotropy profile, $\beta_{\rm cl}(r)$, and
typically this has meant adopting the simplest case of isotropic orbits with
$\beta_{\rm cl}\equiv0$ (see, e.g., Cohen \& Ryzhov 1997). By contrast,
Romanowsky \& Kochanek (2000) apply a technique in which an {\it ad hoc}
functional form of $M_{\rm tot}(r)$ is assumed {\it a priori}, and
the observed $\sigma_p(R)$ is used in conjunction with sophisticated orbit
modelling to constrain both the normalization of the M87/Virgo mass profile
and the behavior of $\beta_{\rm cl}(r)$ in the GCS. This latter procedure is
closer in spirit to ours here, in which we specify both $n_{\rm cl}(r)$ and
$M_{\rm tot}(r)$ {\it in full}, and then use the observed $\sigma_p(R)$
profiles of \S3 to infer something about the anisotropy of the
GC orbits in M87, both globally---for our entire sample---and
separately---for the metal-poor and metal-rich populations.

One way to do this is as an 'inverse problem' (e.g. Merritt \& Oh 1997).
In this approach, the observed function $\sigma_p(R)$ is approximated in 
some model-independent way (e.g., via splines). For an assumed $M_{\rm tot}(r)$, 
the two unknown functions $\sigma_r(r)$ and $\beta(r)$ then follow uniquely 
from the Jeans equation and the deprojection integral.
But given the limited number of velocities in our sample---to say nothing of
our neglect of rotation and our assumption of spherical symmetry---we opt
instead to construct a suite of model $\sigma_p(R)$ profiles by specifying
$\beta_{\rm cl}(r)$ {\it a priori}, and then comparing these models with the data
to gain a broad, qualitative view of the orbital properties of the M87 GCS.
Formally, we solve equation (\ref{eq7}) for $\sigma_r(r)$:
\begin{equation}
\sigma_r^2(r)={1\over{n_{\rm cl}(r)}}\,
\exp\left( -\int {{2\beta_{\rm cl}}\over{r}}\,dr \right)\,
\left[ \int_r^{\infty} n_{\rm cl}\,{{GM_{\rm tot}}\over{x^2}}\,
\exp\left( \int {{2\beta_{\rm cl}}\over{x}}\,dx \right)\, dx \right]\ .
\label{eq8}
\end{equation}
With ${\sigma}_r(r)$ in hand, the projected velocity dispersion profile,
$\sigma_p(R)$, then follows from the standard integral
\begin{equation}
\sigma_p^2(R) =
{2\over{N_{\rm cl}(R)}}\,
\int_R^{\infty} n_{\rm cl} \sigma_r^2(r)
\left(1 - \beta_{\rm cl}\,{{R^2}\over{r^2}} \right)\,
{{r\,dr}\over{\sqrt{r^2 - R^2}}},
\label{eq9}
\end{equation}
where the projected GCS density $N_{\rm cl}(R)$ is given by equation
(\ref{eq3}). We also compute the cumulative, {\it aperture} velocity
dispersion profile, defined to be the rms line-of-sight velocity for all
clusters inside a given radius:
\begin{equation}
\sigma_{\rm ap}^2(\le R) = \left[ \int_0^R N_{\rm cl}(R) \sigma_p^2(R)\,
R\,dR \right]\, \left[\int_0^R N_{\rm cl}(R)\,R\,dR \right]^{-1}\ .
\label{eq10}
\end{equation}
This statistic has the obvious advantage of reduced noise in the measured
dispersion at large radii, although its interpretation can be somewhat more
complicated since it reflects a coarse average (and progressively more so 
as $R$ increases) of a spatially varying quantity.

To proceed in this way, we require accurate
descriptions of: (1) the total gravitating mass profile in M87 (and, as the
projection integrals above make clear, in the Virgo Cluster beyond our last
observed galactocentric radius); and (2) the spatial distribution of the full M87
GCS and of the blue and red clusters, $n_{\rm cl}(r)$, individually.

\begin{figure*}[b]
\centering \leavevmode
\epsfysize=4.0truein
\epsfbox{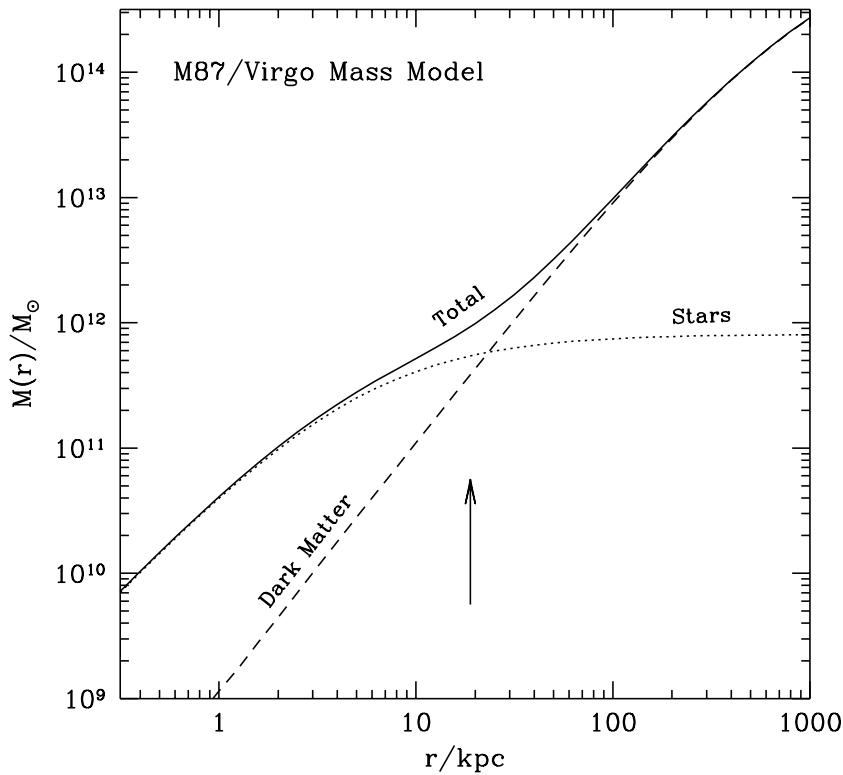}
\caption{Mass model for the central regions of the 
Virgo cluster, adopted from McLaughlin (1999a). The solid curve shows the total mass 
profile: i.e., the combined Virgo cluster dark matter distribution (as traced by 
the x-ray emitting gas) and that of the stellar
component of M87, indicated by the dashed and dotted curves, respectively. 
See \S 4 and equation (11) for details. The vertical arrow shows the
projected radius where the cD enevelope of M87 begins.
\label{fig10}}
\end{figure*}

First, we take $M_{\rm tot}(r)$ directly from McLaughlin (1999a), who has
developed a simple mass model for M87 and Virgo that satisfies all
observational constraints imposed by the optical surface photometry of the
galaxy (as a tracer of the stellar mass density; de Vaucouleurs \& Nieto
1978) by the X-ray surface brightness on $R\la 200$--kpc scales in the core of
the cluster (used under the assumption that the hot intracluster gas around
M87 is in hydrostatic equilibrium with the total mass distribution; Nulsen
\& B\"ohringer 1995), and by the spatial distribution and line-of-sight
velocity dispersion of the early-type (mostly dE) Virgo galaxies out to scales
of $\sim$ 3 Mpc (Binggeli et al.~1985; Girardi et al.~1996). We emphasize that
because this model for
$M_{\rm tot}(r)$ was constructed explicitly without any reference to 
GCS data, {\it it is completely separate from, and independent of, our
radial-velocity observations.} Thus, the total (baryonic and dark) mass
interior to any three-dimensional radius $r$ in Virgo is taken to be,
with normalizations appropriate for a distance of 15 Mpc to the cluster,
\begin{equation}
\begin{array}{rcl}
M_{\rm tot}(r)   & = & M_{\rm stars}(r) + M_{\rm dark}(r) \\
M_{\rm stars}(r) & = & 8.10 \times 10^{11}\,M_\odot\,
            \left[(r/5.1\,{\rm kpc})/(1+r/5.1\,{\rm kpc})\right]^{1.67} \\  
M_{\rm dark}(r)  & = & 7.06 \times 10^{14}\,M_\odot\,
            \left[ \ln(1+r/560\,{\rm kpc}) -
                  (r/560\,{\rm kpc})/(1+r/560\,{\rm kpc})\right].
\end{array}{}
\label{eq11}
\end{equation}
Note that this model attributes all of the dark matter around M87 to the Virgo
Cluster as a whole; {\it all} of the available data are consistent with this
minimalist hypothesis, in which M87 as a galaxy has no dark matter halo of its
own (see McLaughlin 1999a). The functional form of $M_{\rm dark}(r)$ is
that expected for the ``universal'' dark-matter halo of Navarro, Frenk, \&
White (1997), in which the volume density scales as $\rho_{\rm dark}\sim
r^{-1}(r_s+r)^{-2}$, with a fitted value of $r_s=560$ kpc.
The adopted models for $M_{\rm stars}(r)$, $M_{\rm dark}(r)$ and 
$M_{\rm tot}(r)$ are illustrated in Figure~\ref{fig10}. 

\begin{figure*}[t]
\centering \leavevmode
\epsfysize=4.0truein
\epsfbox{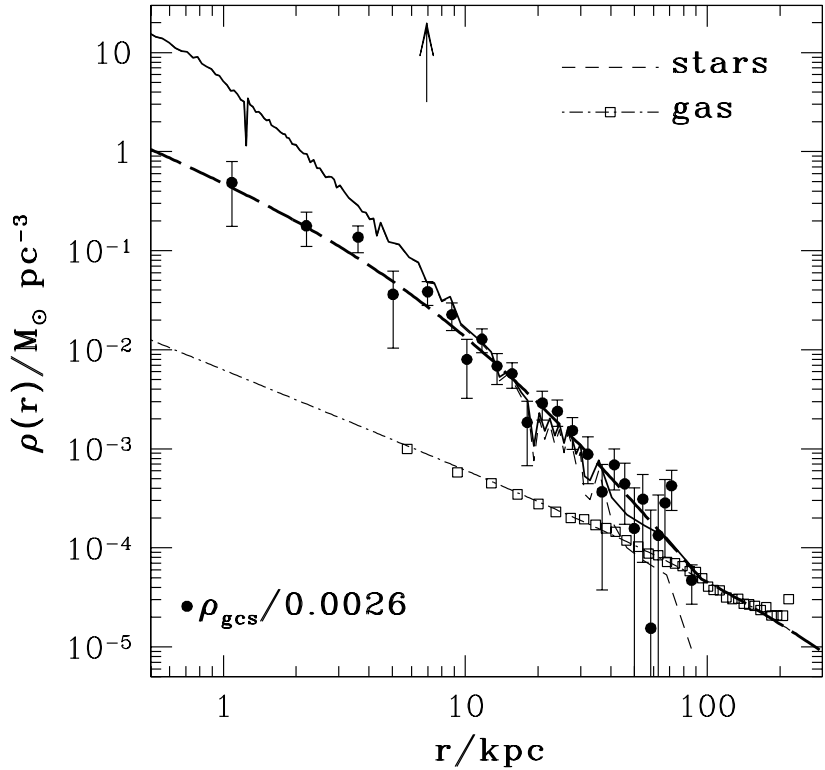}
\caption{Comparison of the (mass) volume density profiles of the
globular cluster system (filled circles and bold dashed line), halo stars
(thin dashed line) and X-ray gas (open squares and dot-dash line) around M87
(after McLaughlin 1999b). Note that ${\rho}_{\rm cl}$ has been scaled upward
by the factor $1/0.0026$ (see text). The solid line is the sum of the stellar and
gas densities, which have been calculated by de-projecting the galaxy surface photometry 
with the algorithm of McLaughlin (1999b). The vertical arrow marks the projected effective 
radius of the galaxy light, $R_{\rm rff} = 96\arcsec\simeq 7$ kpc (de Vaucouleurs \& 
Nieto 1978). Beyond this, it is found that $\rho_{\rm cl} \propto (\rho_{\rm stars}
+ \rho_{\rm gas})$.
\label{fig11}}
\end{figure*}

Next, Figure~\ref{fig11} motivates our favored model for the density profile
of the full M87 GCS. This plot, which is taken nearly in full from McLaughlin
(1999b) and which is described in detail there, shows the volume {\it mass}
densities of the stars in M87 (obtained by applying a spatially constant
mass-to-light ratio to the optical surface photometry of M87), the GCs
around the galaxy (obtained by combining number counts from a number
of different studies and applying a geometrical de-projection algorithm), and
the hot, X-ray emitting gas in the core of Virgo (obtained in a
model-independent way from ROSAT observations, by Nulsen \& B\"ohringer 1995).
The dash-dot line running through the gas datapoints is directly proportional
to the density profile of the dark-matter halo just discussed --- i.e., it shows
that $\rho_{\rm gas}\propto \rho_{\rm dark}\propto (r/560\,{\rm kpc})^{-1}
(1+r/560\,{\rm kpc})^{-2}$ in Virgo (see McLaughlin 1999a).

The GC density data in Figure~\ref{fig11} have been scaled up by
a factor of $1/0.0026 \sim 385$ to illustrate that, beyond a radius of
$r\approx 7$ kpc (the effective radius of the M87 starlight in projection) a
strict proportionality ties the density of the GCS to that of the other
baryons: $\rho_{\rm cl}(r)=0.0026\,[\rho_{\rm stars}(r)+\rho_{\rm gas}(r)]$.
The constant of this proportionality, 0.0026, appears to hold similarly in
nearly 100 other early-type galaxies, and it has been interpreted by McLaughlin
(1999b) as a universal efficiency of GC formation. Thus, for
radii $r\ga 100$ kpc, where there are no direct observations of the GCS
density profile, we assume that $\rho_{\rm cl}(r)\simeq 0.0026\rho_{\rm gas}(r)
\propto\rho_{\rm dark}(r)$. Inside this, we adopt an {\it ad hoc} functional
fit to the GCS density data from McLaughlin (1999b). In total, we posit that
\begin{equation}
\begin{array}{rcll}
n_{\rm tot}(r) & \propto & (r/9.1\,{\rm kpc})^{-1}(1+r/9.1\,{\rm kpc})^{-2}\ ,
               & r < 95\ {\rm kpc} \\
               & \propto & (r/560\,{\rm kpc})^{-1}(1+r/560\,{\rm kpc})^{-2}\ ,
               & r > 95\ {\rm kpc}\ .
\end{array}{}
\label{eq12}
\end{equation}
The normalization of this model, which is drawn as the bold, dashed line
in Figure~\ref{fig11}, is unimportant for our purposes (see eq.~[\ref{eq8}]).
The observed proportionality between the GC and gas density profiles has
some important implications for the debate over whether some or all of the
GCs surrounding M87 are intergalactic in nature (White 1987; West et~al. 1995;
Harris et~al. 1998). We shall return to this issue in \S 5.7.

\begin{figure*}[t]
\centering \leavevmode
\epsfysize=4.0truein
\epsfbox{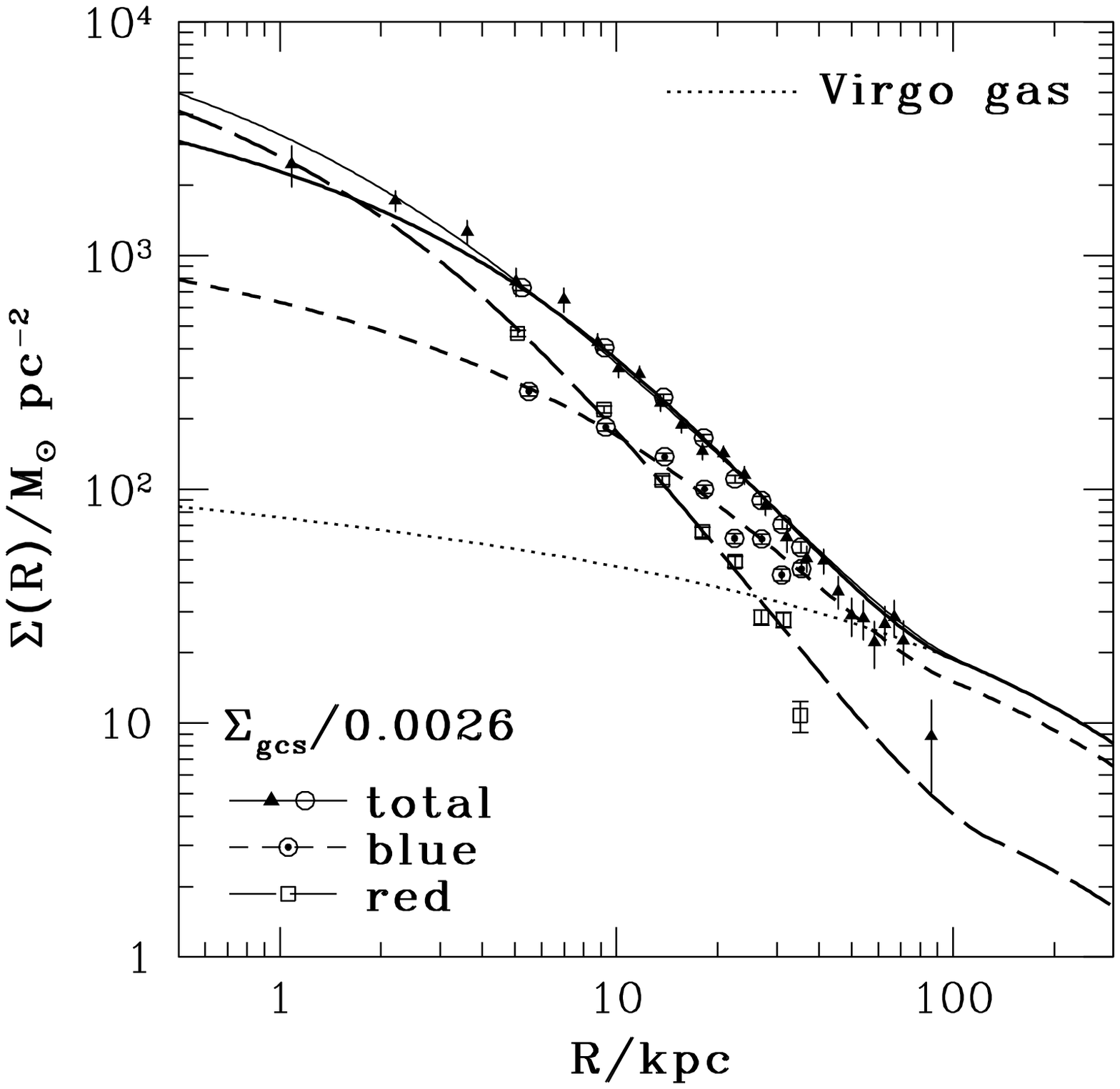}
\caption{Surface density profiles, ${\Sigma}(R)$, for globular
clusters associated with M87. Triangles show the surface density profile of
McLaughlin (1999b) for the full globular cluster system, while open circles show 
the profile found here using the Washington photometry described in Paper I. 
Circled points and open squares indicate the respective surface density profiles 
of metal-poor (blue) and metal-rich (red) globular clusters. The bold curves 
correspond to projections of the models in equations (12)--(14) in the text; 
see the discussion there for details.
\label{fig12}}
\end{figure*}

Figure~\ref{fig12} shows the projection of this density model as a
bold solid line. (The mass surface density $\Sigma$ is related to $\rho$ in
the same way as the projected number density $N$ is related to $n$.) The
small triangles are the observed surface densities, again taken from
McLaughlin (1999b), that were de-projected to provide the volume densities in
Figure~\ref{fig11}. We have now also used our new CT$_1$ photometry to
measure the surface density profiles, not only for the complete GC
sample, but also for the metal-poor and metal-rich samples separately.
To do so, our initial catalog of point sources was first trimmed to exclude
all objects having T$_1$ $\le$ 21, C$-$T$_1$ $\le$ 0.8, and C$-$T$_1$ $\ge$ 2.35,
leaving us with a sample of 2130 candidate GCs. The adopted
limiting magnitude was chosen to ensure that our photometric catalog is
complete over the full range in galactocentric radius (see Geisler, Lee \&
Kim 2001). A background surface density of 0.8 clusters arcmin$^{-2}$ was then
subtracted to give the surface density profile for the entire cluster
population. This background level was chosen based on a comparison of our
surface density profile with the calibrated surface density profile of
McLaughlin (1999b), and agrees to $\simeq$ 5\% with that predicted by the
Galactic star count model of Bahcall \& Soneira (1981). The final,
blue-plus-red profile is shown as the open circles in Figure~\ref{fig12},
where it has been scaled vertically to match the counts from McLaughlin
(1999b) and shows good agreement with that independent measurement of
$\Sigma_{\rm cl}(R)$.

\begin{figure*}[t]
\centering \leavevmode
\epsfysize=4.0truein
\epsfbox{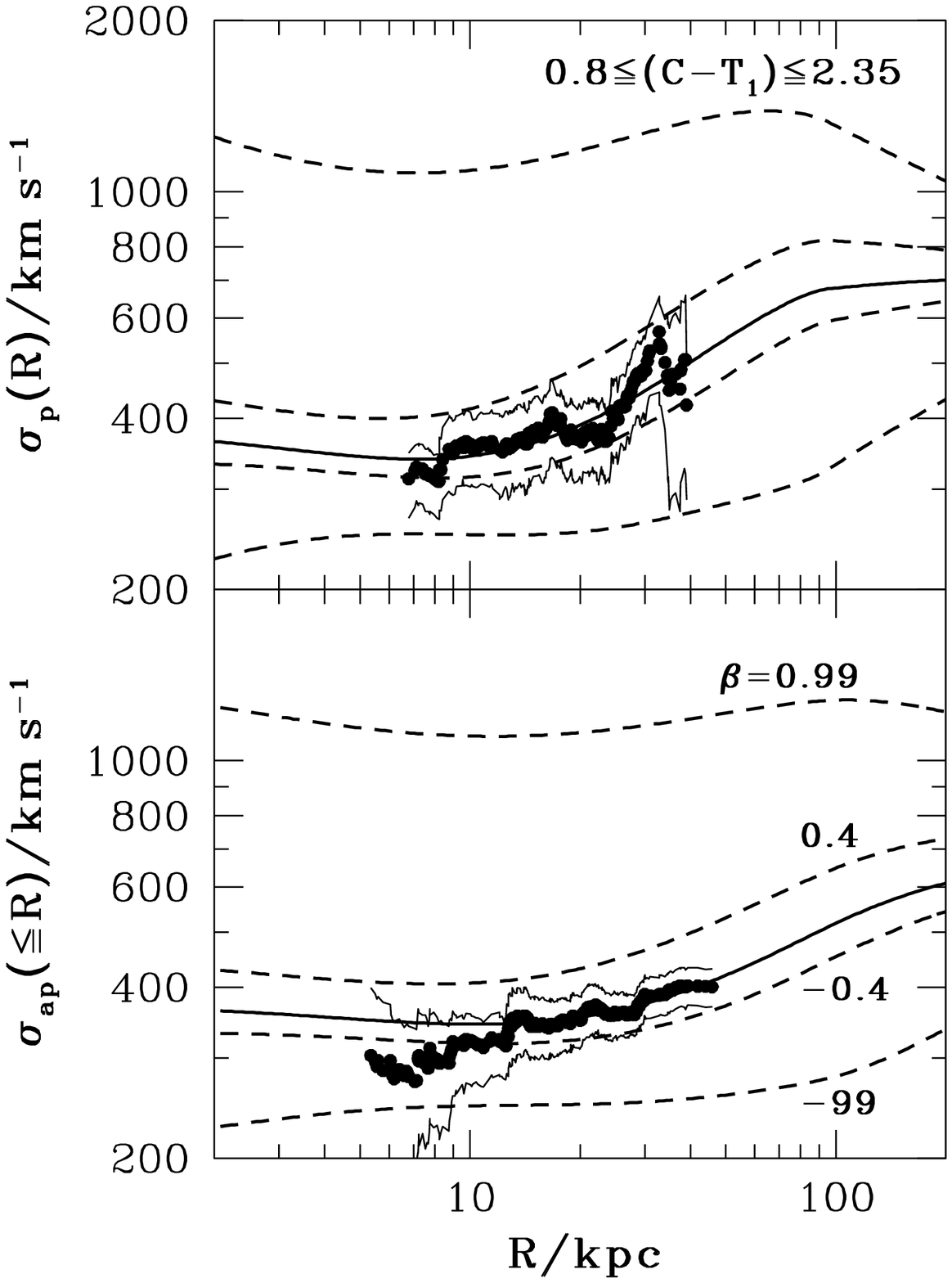}
\caption{{\it Upper Panel}: Velocity dispersion profile for
the complete sample of globular clusters. Points represent biweight
estimates for the line-of-sight velocity dispersion in radial bins of width
$\Delta R = 120\arcsec\simeq 8.7$ kpc; thin solid curves show bootstrap
estimates of the 90\% confidence limits on these measurements. The thick solid
line shows the {\it predicted} velocity dispersion profile for a system of
test particles that are embedded in the potential defined by equation (11),
follow a density profile given by equation (12) (the bold, dashed line in
Figure~\ref{fig11}), and have an isotropic velocity ellipsoid ($\beta_{\rm cl}
\equiv 0$). From top to bottom, the four dashed curves show the predicted
velocity dispersion profiles for  $\beta_{\rm cl} \equiv +0.99$ (a strong
radial bias), $\beta_{\rm cl} \equiv +0.4$ (a moderate radial bias),
$\beta_{\rm cl} \equiv -0.4$ (a moderate tangential bias) and $\beta_{\rm cl}
\equiv -99$ (a strong tangential bias). {\it Lower Panel}: Aperture velocity
dispersion profile for the complete sample of globular clusters, with
bootstrap estimates of the 90\% confidence limits illustrated by thin solid
lines. The heavy solid curve shows the {\it predicted} aperture velocity
dispersion profiles for the case of isotropic orbits, while the four dashed
curves show model predictions with the same velocity anisotropies as in the
above panel.
\label{fig13}}
\end{figure*}

\begin{figure*}[t]
\centering \leavevmode
\epsfysize=4.0truein
\epsfbox{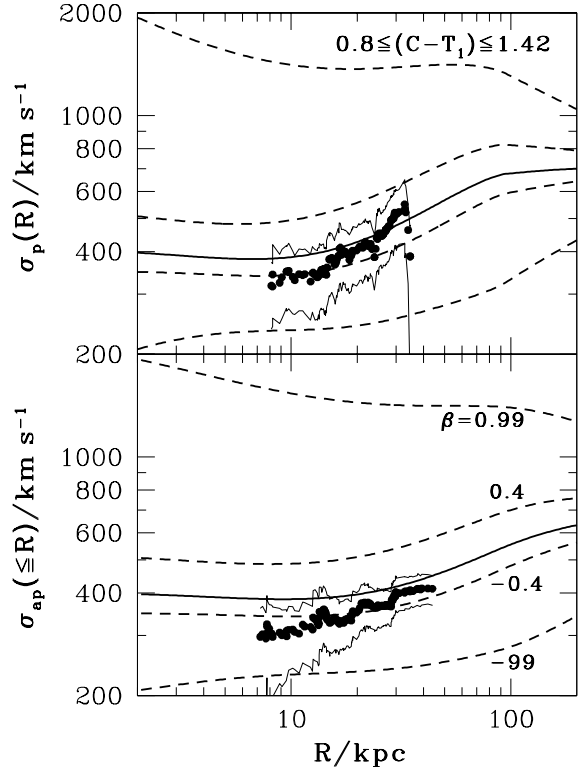}
\caption{Same as Figure~\ref{fig13}, except for the sample of
metal-poor (blue) globular clusters. Note that the aperture velocity
dispersion at small radii in the lower panel falls below that expected in
the case of isotropic orbits, at the 90\% confidence level. The model which
best matches the measured profile has $\beta_{\rm cl} \approx -0.4$,
corresponding to a {\it tangential} bias of $\sigma_{\theta} \approx
1.2{\sigma}_{\rm r}$.
\label{fig14}}
\end{figure*}

\begin{figure*}[b]
\centering \leavevmode
\epsfysize=4.0truein
\epsfbox{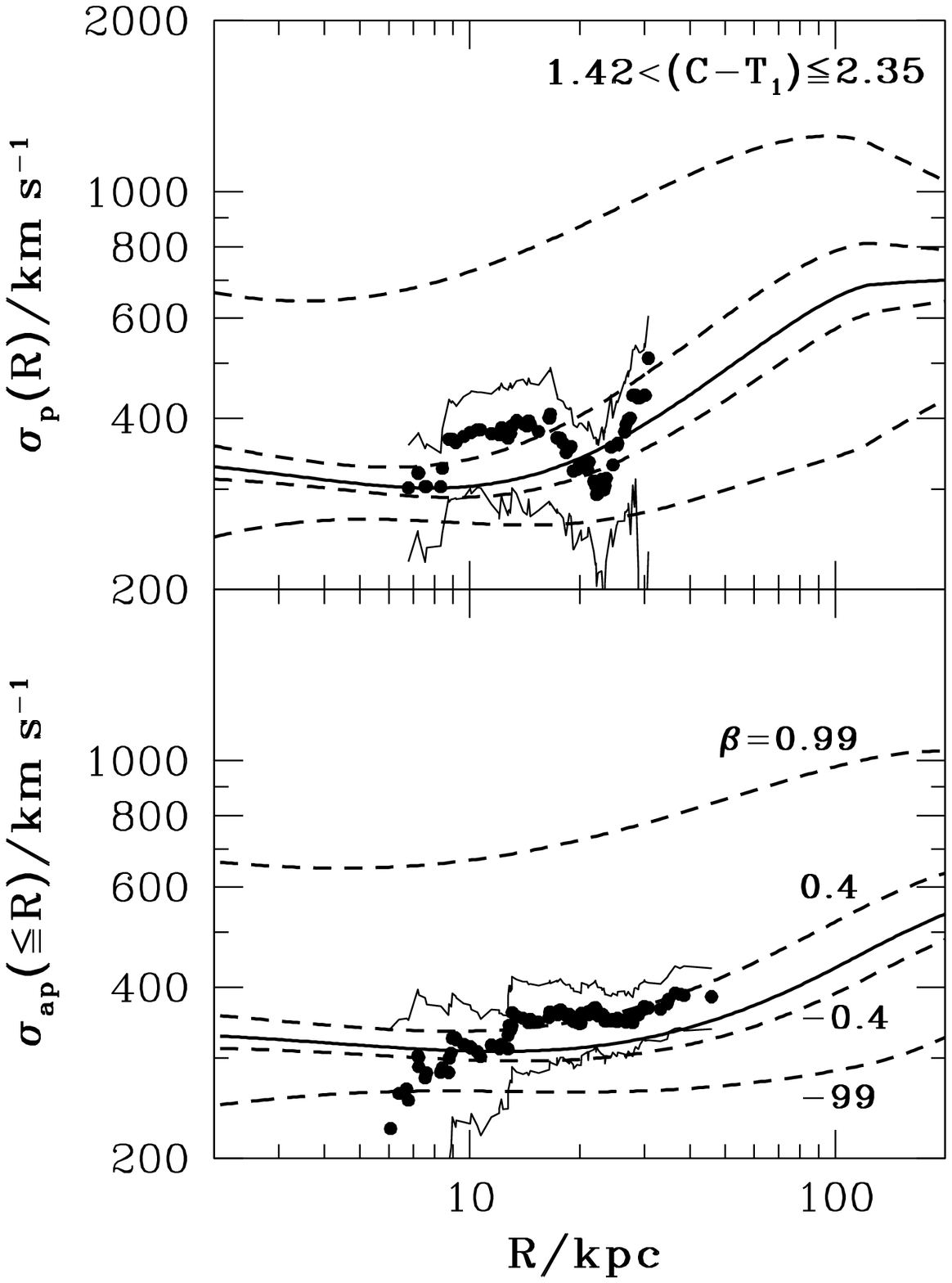}
\caption{Same as Figure~\ref{fig13},  except for the sample of
metal-rich (red) globular clusters. In this case, the aperture velocity
dispersion profile at {\it large} radii in the lower panel lies above that
expected for an isotropic velocity ellipsoid. A model which reasonably
matches the measured profile has $\beta_{\rm cl} \approx +0.4$, corresponding
to a {\it radial} bias of $\sigma_{\theta} \approx 0.8{\sigma}_{\rm r}$.
\label{fig15}}
\end{figure*}

For the separate metal-rich and metal-poor samples, we have taken the
background surface densities to be exactly half that of the total background 
in our CT$_1$ sample. Though somewhat arbitrary, these choices of backgrounds
are consistent with the color distribution of foreground stars predicted
by Ratnatunga \& Bahcall (1985) for three Galactic GCs having Galactic latitudes and
longitudes similar to M87. This approach is obviously inferior to measuring
surface densities directly from background fields, but the precise choice of
backgrounds has relatively little effect on the final surface density 
profiles, except in the outermost bins. The resulting distributions for the
metal-rich and metal-poor cluster populations are indicated by the open squares
and circled points in Figure~\ref{fig12}. This shows quite clearly that
the metal-poor clusters in M87 follow a much more shallow radial profile
than their metal-rich counterparts, consistent with earlier findings based on 
{\sl HST} imaging in the central regions of the galaxy (e.g., 
Neilsen, Tsvetanov \& Ford 1999).  There is no reasonable choice of
background that can alter this basic conclusion.
Even if we were to make the highly
implausible assumption that the background consists {\it entirely}
of sources having colors of C$-$T$_1$ $\le$ 1.42, this qualitative contrast 
between the blue and red GCSs would remain.

A more detailed discussion of the spatial distribution of the M87 GCS
will be presented in Geisler, Lee \& Kim (2001); our purposes
only require estimations of the overall shapes of $n_{\rm blue}(r)$ and
$n_{\rm red}(r)$. To obtain these, we project models similar in form to that
in equation (\ref{eq12}) and fit to the blue- and red-GCS data points in 
Figure~\ref{fig12}. We obtain
\begin{equation}
\begin{array}{rcll}
n_{\rm blue}(r) & \propto
                & (r/20.5\,{\rm kpc})^{-1}(1+r/20.5\,{\rm kpc})^{-2}\ ,
                & r < 95\ {\rm kpc} \\
                & \propto
                & (r/560\,{\rm kpc})^{-1}(1+r/560\,{\rm kpc})^{-2}\ ,
                & r > 95\ {\rm kpc}\ ,
\end{array}{}
\label{eq13}
\end{equation}
and
\begin{equation}
\begin{array}{rcll}
n_{\rm red}(r) & \propto & (r/3.3\,{\rm kpc})^{-1}(1+r/3.3\,{\rm kpc})^{-2}\ ,
               & r < 125\ {\rm kpc} \\
               & \propto & (r/560\,{\rm kpc})^{-1}(1+r/560\,{\rm kpc})^{-2}\ ,
               & r > 125\ {\rm kpc}\ .
\end{array}{}
\label{eq14}
\end{equation}
The projections of these best-fit models are drawn as the bold, broken lines in
Figure~\ref{fig12}. Their absolute normalization is again unimportant for us,
although their {\it relative} normalization does matter. This has been set so
that $n_{\rm blue}/n_{\rm red}\rightarrow 4$, independent of radius in the
limit $r \rightarrow \infty$. The projected sum of these red and blue models
is shown in Figure~\ref{fig12} as the thin solid line, which is reasonably
compatible with the completely independent projection of $n_{\rm tot}(r)$
from equation (\ref{eq12}).

With specifications for the Virgo mass profile $M_{\rm tot}(r)$ and for
the density profiles $n_{\rm cl}(r)$ of the total, blue, and red GCSs thus
in hand, we solve equations (\ref{eq8})--(\ref{eq10}) for five different
{\it spatially constant} values of the GCS orbital anisotropy 
$\beta_{\rm cl}$\footnotemark
\footnotetext{It is worth emphasizing that these values of $\beta_{\rm cl}$
are best interpreted as {\it density-weighted averages} due to our
assumption of spatially constant anisotropies.}
\begin{equation}
\beta_{\rm cl} \equiv \left\{ 
\begin{array}{rl}
+0.99   & \quad (\sigma_\theta^2/\sigma_r^2=0.01, {\rm radial\ bias}) \\
+0.4    & \quad (\sigma_\theta^2/\sigma_r^2=0.6, {\rm radial\ bias}) \\
0       & \quad (\sigma_\theta=\sigma_\phi=\sigma_r, {\rm isotropic}) \\
-0.4    & \quad (\sigma_\theta^2/\sigma_r^2=1.4, {\rm tangential\ bias}) \\
-99     & \quad (\sigma_\theta^2/\sigma_r^2=100, {\rm tangential\ bias})\ .
\end{array}{}
\right.
\label{eq15}
\end{equation}
Figures~\ref{fig13}--\ref{fig15} show the results of this exercise applied
to our three GCS samples. In each figure, the top panel shows the observed
$\sigma_p(R)$ profile computed by smoothing the data with a radial kernel
of fixed width $120\arcsec\simeq8.2$ kpc (rather than the $90\arcsec$ used for 
Figure~\ref{fig6}). Also shown are the 90\% confidence bands, obtained by
bootstrapping as described in \S3. The bottom panels show the aperture
velocity dispersion and bootstrap estimates of the 90\% confidence bands for
the cluster color ranges indicated. Overlaid on the data in every case are the
model curves predicted by substituting equations (\ref{eq11})--(\ref{eq15}) in
equations (\ref{eq8})--(\ref{eq10}). We emphasize that these theoretical velocity
dispersion profiles have not been fit to the data: with $M_{\rm tot}(r)$,
$n_{\rm cl}(r)$, and $\beta_{\rm cl}$ all specified {\it a priori}, there are no
additional parameters to be adjusted. Note that the rising velocity 
dispersion profile seen in Figure~\ref{fig6} is once again evident
here --- not only in the {\it measured} $\sigma_p(R)$ and $\sigma_{\rm ap}(R)$ 
profiles, but also in the {\it models}. This predicted rise in $\sigma_p(R)$ 
and $\sigma_{\rm ap}(R)$ comes about because the background mass density 
profile given by equation~\ref{eq11}, and shown in Figure~\ref{fig10},
is shallower at these radii than that of an isothermal sphere.

It is clear from Figure~\ref{fig13} that the M87 GCS
{\it as a whole} has an almost perfectly isotropic velocity ellipsoid. Models
with even modest amounts of anisotropy, $\beta_{\rm cl} = \pm 0.4$, produce
noticeably inferior matches to observed velocity dispersions, while strongly
tangentially biased orbits ($\beta_{\rm cl}=-99$) can be ruled out with
$>90\%$ confidence and very radial orbits ($\beta_{\rm cl}=0.99$) are
altogether out of the question. However, if the cluster sample is 
divided on the basis of metallicity, some interesting differences emerge.
Figure~\ref{fig14} suggests a modest tangential bias in the
metal-poor clusters, with $\beta_{\rm cl}\sim -0.4$ at small radii
(corresponding to $\sigma_\theta \sim 1.2 \sigma_r$), perhaps tending towards
isotropy at larger radii $\sim$20--30 kpc. The opposite result is seen for the
metal-rich GCs, which have a slight {\it radial} bias of roughly
the same magnitude: Figure~\ref{fig15} suggests ${\beta}_{\rm cl}
\sim +0.4$, and thus $\sigma_{\theta} \sim 0.8{\sigma}_r$. This is just what
is needed, of course, to balance the anisotropy in the blue GCS and produce
the isotropy indicated for the combined sample in Figure~\ref{fig13}.
Qualitatively speaking, these contrasting anisotropies are also what might 
have been expected {\it a priori}, given the fact that the blue GCS has a much shallower
density profile than the red subsystem, even though the two populations are
embedded in the same gravitational potential and show no significant
differences in their projected kinematics (see Figure~\ref{fig6}).

Romanowsky \& Kochanek (2000) have derived a $\beta_{\rm cl}(r)$ profile for the
{\it total} GCS --- i.e., they did not separate their velocity sample on the basis
of metallicity. At small radii, the implied $\beta_{\rm cl}$ is markedly negative, 
thereby implying a substantial tangential bias for the GCS. However, this claim 
ultimately rests on their assumption that the background potential of M87/Virgo is
distributed as a singular isothermal sphere, $M_{\rm tot}(r)\propto r$, whereas
the true $M_{\rm tot}(r)$ in the core of Virgo is well
constrained (principally by X-ray data; see Nulsen \& B\"ohringer 1995 and
McLaughlin 1999a; see also Figure~\ref{fig10}) to grow substantially 
more rapidly than this with increasing
radius. Said another way, the density distribution of the gravitating mass in
the core of Virgo is significantly shallower than the ${\rho} \propto r^{-2}$
distribution of an isothermal sphere, and therefore the density distribution 
of the central GCS does not require so large a tangential bias as Romanowsky 
\& Kochanek (2000) found necessary to satisfy the Jeans equation.

\begin{figure*}[t]
\centering \leavevmode
\epsfysize=4.25truein
\epsfbox{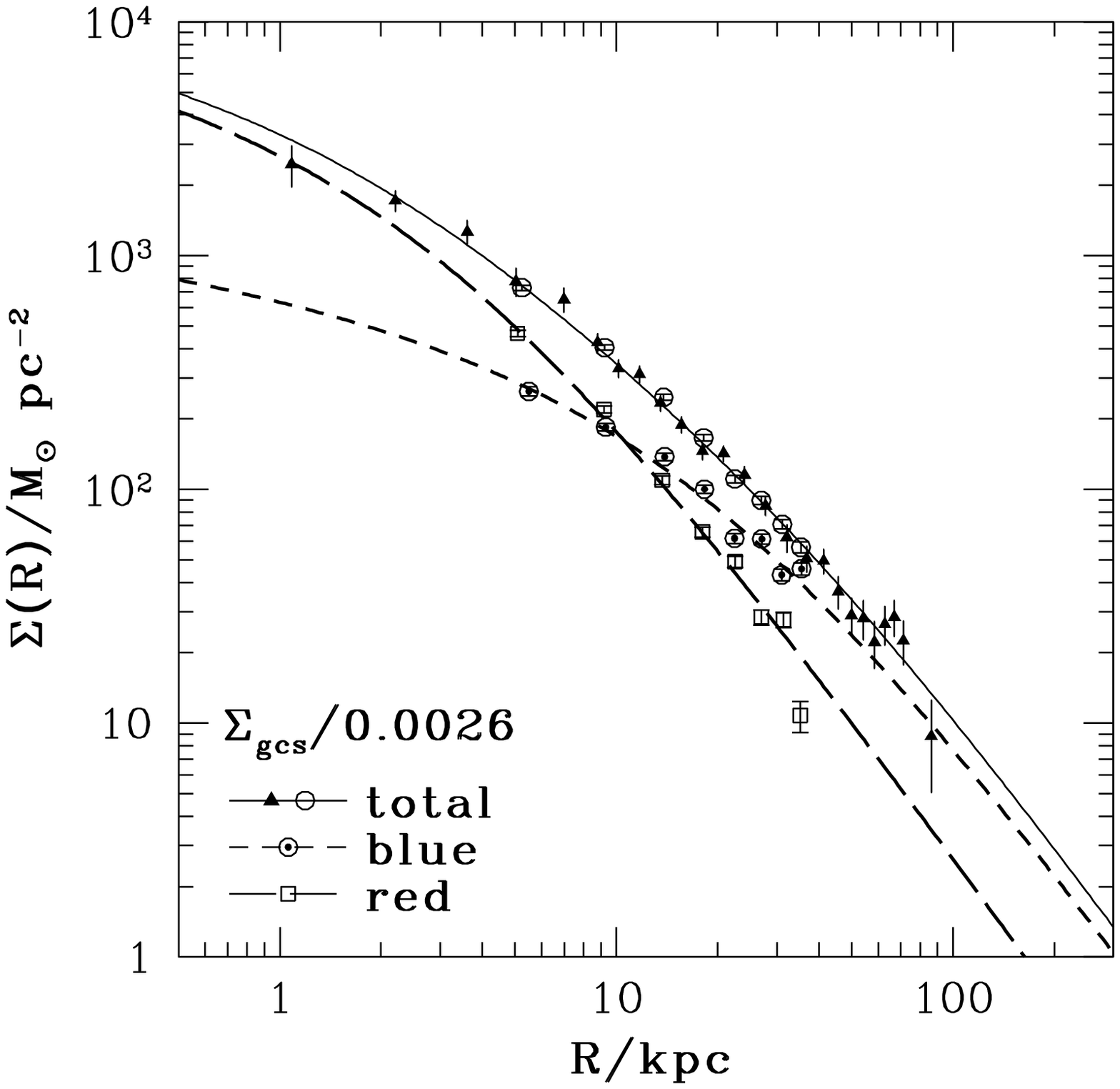}
\caption{Alternate extrapolation of the red and blue globular 
cluster system density profiles to larger, unobserved radii (compare with 
Figure~\ref{fig12}). Thin, solid line is the sum of the red
and blue fits (shown as thick lines) and is in good agreement with the
independently observed density profile of the full globular cluster system.
\label{fig16}}
\end{figure*}

\begin{figure*}[b]
\centering \leavevmode
\epsfysize=4.0truein
\epsfbox{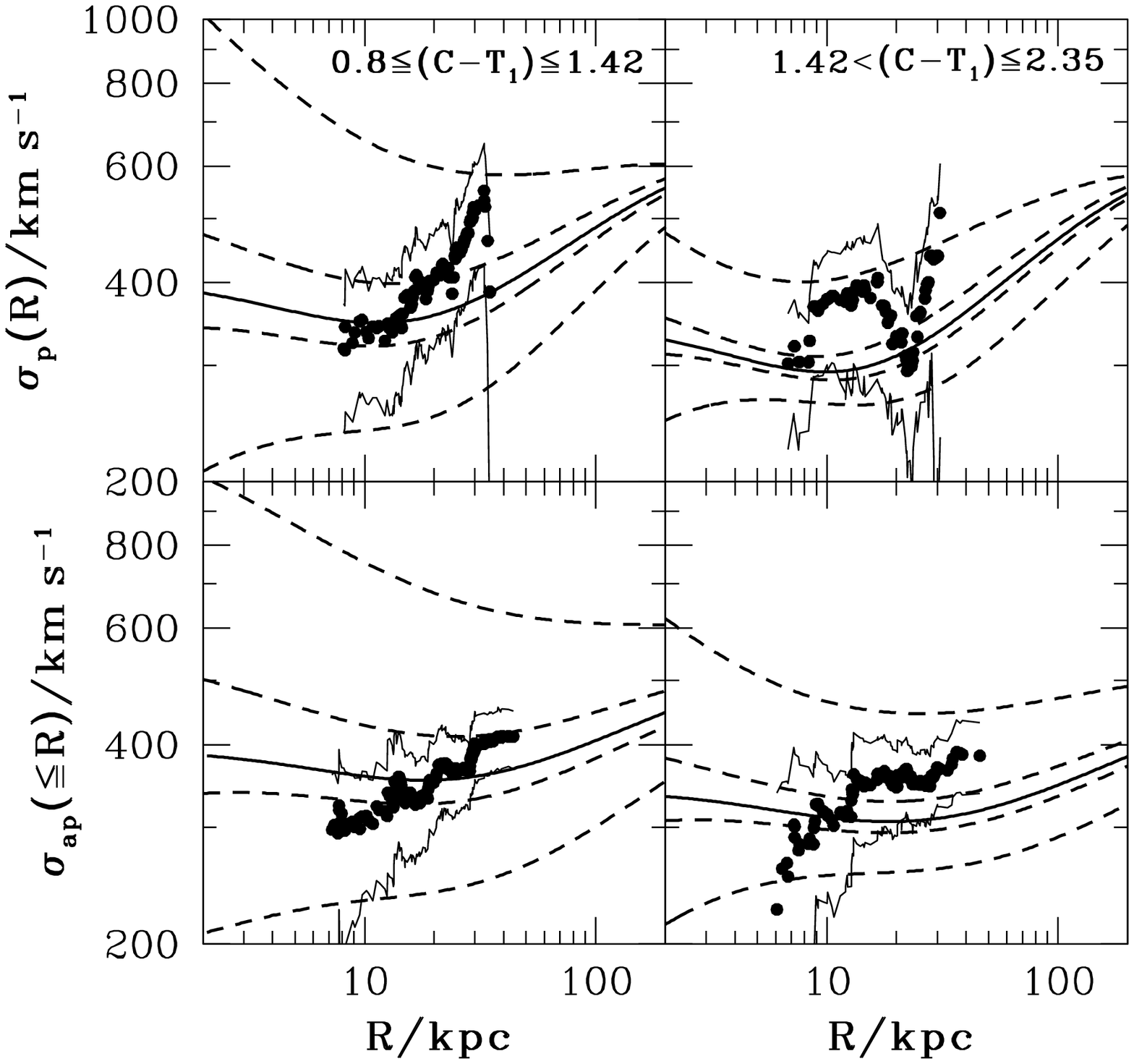}
\caption{Model vs.~observed dispersion profiles, shown
separately for the blue and red subsets of the M87 globular cluster system, assuming the 
alternate density profiles shown in Figure~\ref{fig16} and described by equations (16) and (17).
Since the mass profile of M87/Virgo is fixed in this analysis, the data in
this case prefer models with stronger radial bias in the globular cluster system orbits at large
galactocentric radii. Curves shown have constant $\beta=0.99$, 0.4, 0, $-0.4$
and $-99$, as in Figures~\ref{fig13}--\ref{fig15}.)
\label{fig17}}
\end{figure*}

Finally, we consider the effect on our results if our adopted extrapolation
of the GCS density profiles in Figure~\ref{fig12} and equations
(\ref{eq12})--(\ref{eq14}) is somehow in error. It seems unlikely that the
GC distribution could become shallower than the hot intracluster
gas (and the dark matter) in Virgo, and thus Figures~\ref{fig16} and
\ref{fig17} explore the implications of an $n_{\rm cl}(r)$ that drops off
very steeply towards large radii. Rather than requiring $n_{\rm cl}(r) \propto
\rho_{\rm gas}(r)$ at the unobserved $r\ga 100$ kpc, then, we take at face
value simple fits to our CT$_1$ density profiles for $5\la R\la 30$ kpc and
assume
\begin{equation}
n_{\rm blue}(r) \propto (r/21.3\,{\rm kpc})^{-1}(1+r/21.3\,{\rm kpc})^{-2}\ ,
\qquad {\rm all\ radii} 
\label{eq16}
\end{equation}
and
\begin{equation}
n_{\rm red}(r) \propto (r/3.3\,{\rm kpc})^{-1}(1+r/3.3\,{\rm kpc})^{-2}\ ,
\qquad {\rm all\ radii.}\ 
\label{eq17}
\end{equation}
The bold, broken lines in Figure~\ref{fig16} compares the projection of these
functions to the same data considered in Figure~\ref{fig12}. The thin, solid
line, which shows the sum of the red and blue projected densities, is again
in good agreement with the independently observed density distribution of the
full GCS.

We use these new density profiles with the same $M_{\rm tot}(r)$ from
equation (\ref{eq11}) and the same five constant $\beta_{\rm cl}$ values from
equation (\ref{eq15}) to compute new models for the $\sigma_p(R)$ and
$\sigma_{\rm ap}(\le R)$ of the metal-poor and the metal-rich components of
the GCS. Figure~\ref{fig17} compares these to the data,
which again have been smoothed with a $120\arcsec$-wide sliding radial bin
and are shown with bootstrap estimates of their 90\% confidence bands.
Because the large-$r$ density gradient of both cluster subsystems has been
made steeper by hypothesis without changing any other aspect of the dynamical
models, the data would appear now to suggest
a move towards substantial radial bias in orbits of both blue and red clusters
at the largest radii covered here. We stress, however, that we regard this
particular scenario as a very extreme possibility, as the GCS density
profiles postulated earlier are certainly closer to the truth. For instance, if 
$n_{\rm cl}(r)$ indeed behaves as in Figure~\ref{fig16} rather than as in 
Figure~\ref{fig12}, then we must conclude that the GC-to-baryon mass ratio of 0.0026
is only {\it coincidentally} realized, and then over a limited range of radius,
in M87 --- even though this ``universal" ratio matches precisely what is found 
by McLaughlin (1999b) for nearly 100 other early-type galaxies. 
In any case, Figure~\ref{fig17} also shows that the GC
volume density profiles {\it cannot fall off more rapidly than
roughly $n_{\rm cl}\propto r^{-3}$}, since this gradient already requires
the clusters at large $r$ to have nearly purely radial velocity ellipsoids
in order to reproduce the observed line-of-sight velocity dispersion.

\section{Discussion}

\subsection{Results for other Ellipticals from Globular Clusters}

Before turning our attention to the various models which have been proposed
for the formation of GCSs and their host galaxies, we pause first to 
compare our findings for the GCS of M87 with those of other elliptical 
galaxies. Perhaps the most natural comparison is that of M49, the most luminous
member of the Virgo cluster, whose GCS has been studied recently by 
Sharples et~al. (1998) and Zepf et~al. (2000). From a sample of 144 GCs 
with measured velocities, Zepf et~al. (2000) report that the metal-poor and 
metal-rich GCSs have different dynamical properties: i.e.,
$v_{\rm rot}/\sigma \sim 0.3$ and $0.1$, respectively (with $v_{\rm rot}/\sigma
< 0.56~{\rm and}~0.34$ at 99\% confidence). These 
estimates are to be compared with the value of $v_{\rm rot}/\sigma \sim 
0.45$ which we measure for both components in M87. Hui et~al. (1995)
have analysed the radial velocities of 433 planetary nebulae (PNe)
and 62 globular clusters in NGC 5128, and conclude
that the metal-rich GCs in that galaxy show roughly the same 
dynamical properties as do the PNe, $v_{\rm rot}/\sigma \sim 0.3$, 
whereas the metal-poor GCs show no evidence of rotation. Meanwhile,
Kissler-Patig et~al. (1999) find some marginal evidence for rotation
among the outermost GCs in NGC 1399, the central elliptical in Fornax, 
but due to the limited size of their radial velocity sample,
were unable to ascertain if the chemically distinct GC populations 
differ in their dynamical properties. Thus, at this early stage, 
no obvious pattern in the dynamical properties of the GCSs of elliptical 
galaxies has yet emerged.

\subsection{Results for other Ellipticals from Planetary Nebulae}

Radial velocity surveys of PNe represent an alternative, and complementary, 
method of studying the halo dynamics of giant
elliptical galaxies. While there have been no dynamical studies 
of the PNe population of M87 itself, some
comparisons are possible for the three luminous early-type galaxies having
large numbers of measured PNe radial velocities: NGC 5128 (Hui et
al.~1995), NGC 4406 (Arnaboldi et al.~1996) and NGC 1316 (Arnaboldi et
al.~1998). The PNe in all three galaxies show evidence for
significant rotation, $v_{\rm rot}/\sigma \sim 0.5$-1, in their outer regions. 
In this respect, these galaxies resemble M87 which has 
$\langle{(\Omega R)/\sigma}\rangle \sim 0.5$ from its GCS, with some evidence
for slightly lower values in the inner regions of the galaxy and an outward
increase in ${(\Omega R)/\sigma}$. Moreover, integrated-light spectroscopy in 
the inner regions of these galaxies reveals little or no rotation in each case,
similar to the situation for luminous elliptical galaxies, and for 
M87 in particular (e.g., $v_{\rm rot,*}/\sigma_* \lae 0.1$ according to Davies
\& Birkinshaw 1988). On the other hand, none of these galaxies
has the generally round and smooth photometric appearance of M87: NGC 5128
and NGC 1316 are clear examples of recent gas-rich mergers or accretions (e.g.,
Schweizer 1980), and NGC 4406 is probably best classified as an S0 galaxy
(Arnaboldi et al.~1996). While firm conclusions must await the measurement of
radial velocities for PNe in the vicinity of M87, 
our results for the M87 GCS appear consistent with the claim
of Arnaboldi et al.~(1998) that giant elliptical galaxies, by virtue of their
relatively rapid rotation at large radii, may contain roughly as much angular 
momentum per unit mass as do spiral galaxies {\it assuming that the GC and PNe 
kinematics are representative of the kinematics of the underlying galaxy mass.}

\subsection{The GCS Velocity Ellipsoid at Small Galactocentric Radii}

It has been known for some time that the surface density profile of the
GCs surrounding M87 exhibits a ``core" of radius $R_c \simeq$ 4-5 kpc
(Lauer \& Kormendy 1985; McLaughlin 1995). The possibility that this
core developed as a result of GC destruction through tidal shocks
from a central bulge or black-hole, or from an underlying
triaxial potential, has been explored by many investigators
(e.g., Ostriker, Binney \& Saha 1989; Pesce, Capuzzo-Dolcetta \& 
Vietri 1992). These authors also noted that such processes are
expected to lead to a preponderance of tube orbits among the 
surviving GCs relative to the initial sample. Moreover, in the specific
case of the Galactic GCS, Murali \& Weinberg (1997) have shown that
GCs on eccentric orbits will undergo the most rapid relaxation and
evaporation due to enhanced tidal heating; the preferential destruction of
such clusters will therefore lead to an orbital distribution which
becomes more tangentially biased with time. 
Thus, depending of course on the {\it initial} shape of 
the GCS velocity ellipsoid, the tangential bias
which we find for the GCs at small galactocentric radii --- especially
within an effective radius (see Figures~\ref{fig13}--\ref{fig15}) --- 
may provide evidence that these dynamical processes have played a role 
in shaping the presently observed GCS. 

\subsection{Cooling Flow Models}

Fabian, Nulsen \& Canizares (1984) were the first to propose that the
populous GCSs associated with some cD galaxies might be the signature of 
GC formation through cooling flows from their X-ray halos. A number of difficulties
with this scenario have been raised over the years: e.g., most ``high-S$_N$" and/or 
cooling-flow galaxies show no evidence for young clusters, the GCs
are generally more metal-deficient than the intracluster gas, and 
it is unclear how the cooling gas might condense into clumps of mass 
$\sim10^5 M_\odot$. For M87, which is both an incipient cD galaxy and the
prototypical ``high--S$_N$" system, we may now add another difficulty to this
list: as described in \S 4, the orbits of the surrounding GCs are closely isotropic, 
{\it not the radial orbits expected if the GCs have condensed out of the
infalling gas.} As Figures~\ref{fig13}--\ref{fig15} demonstrate, purely radial 
orbits are ruled out at very high significance by our observations. Thus, it 
appears highly unlikely that a large fraction of M87's GCs could have formed 
by mass drop-out from a cooling flow.

\subsection{Monolithic Collapse Models}

Can the relatively rapid rotation of the M87 GCS be accommodated within the
context of formation models which argue for the monolithic collapse of a single 
protogalactic gas cloud? Like those of most other giant elliptical galaxies, the 
M87 GCS is disjoint in terms of its chemical properties (i.e., the GC metallicity 
distribution function is decidedly bimodal; Whitmore et~al. 1995; Geisler et~al. 
2001; see also Figure~\ref{fig1}), and this observation 
alone provides a compelling argument against the monolithic collapse picture.
Figures~\ref{fig7}--\ref{fig9} may provide additional evidence against this
model: both the metal-poor and metal-rich GC subsamples exhibit significant
rotation, with ${\Omega}R \simeq 170$ km s$^{-1}$ and 
$\langle{({\Omega R})/{{\sigma}_p}}\rangle$ = 0.45. In monolithic collapse
scenarios, the angular momentum of galaxies can only arise from tidal torques 
from companions (Peebles 1969). A common measure of a galaxy's angular
momentum content is
the dimensionless spin parameter, ${\Lambda} = J|E|^{1/2}G^{-1}M^{-1/2}$, where
$J$, $E$ and $M$ are the angular momentum, binding energy and
mass of the collapsed galaxy. Fall (1979) showed that, for an elliptical
galaxy in gravitational equilibrium,
${\Lambda} \sim 0.3\langle{({\Omega R})/{{\sigma}_p}}\rangle$.
Thus, the inferred spin parameter of the M87 GCS is ${\Lambda} \simeq 0.15$ ---
lower than, but still consistent with, the value of ${\Lambda} \simeq 0.18$
found by Kissler-Patig \& Gebhardt (1998) from a smaller sample of GCs. This
spin parameter is roughly twice that expected from analytical and numerical
simulations of the collapse of a single, isolated protogalactic cloud 
(Peebles 1969; Efstathiou \& Jones 1979). It is, however, worth bearing in mind
that M87 is by no means an isolated system. Perhaps a more severe obstacle
for such models is the abrupt shift in the orientation of the rotation
axis for the metal-poor GC subsample at $R \simeq$ 16-18 kpc. If real, this
feature would be difficult to explain in models where M87 and its GCS formed
via the collapse and spin-up of a single protogalactic cloud.

\subsection{Merger and Accretion Models}

Two variations on the general theme of mergers and accretions have been
proposed for the formation of globular cluster systems:
(1) {\it major mergers} involving a pair of gas-rich disk galaxies 
(each containing only metal-poor GCs) which results in the formation of new, 
metal-rich GCs (Ashman \& Zepf 1992); and (2) {\it 
hierarchical growth} of a pre-existing ``seed" galaxy (containing 
its own metal-rich GCS) through the accretion of 
numerous smaller galaxies or protogalactic fragments which harbor
predominantly metal-poor GCs (C\^ot\'e,
Marzke \& West 1998). We discuss the viability of these 
two models in light of the dynamical evidence presented above.

We begin by reiterating the well established result that mergers
of {\it purely stellar disks} are a highly unlikely mechanism for
the production of giant elliptical galaxies. Both N-body 
simulations ($e.g.$, Barnes 1992; Hernquist 1992) and simple 
arguments regarding central phase-space densities (Carlberg 1986) have 
demonstrated that the end-products of such mergers have cores 
which are far more diffuse than the centers of giant elliptical 
galaxies. In addition, luminous elliptical galaxies typically have Hubble
Types of E2 (Franx, Illingworth \& de Zeeuw 1991), whereas the
end-products of pairs of stellar disks are frequently more elongated,
with Hubble Types of E3-E7 (Hernquist 1992).\footnotemark
\footnotetext{And M87 itself has a Hubble Type of E2 (Van der Marel 1991), 
though the ellipticity does increase outward, reaching a maximum of E4 in 
the cD envelope (see Figure~\ref{fig18}).} Finally, the merger
end-products often show large misalignments between their rotation
and minor axes (Barnes 1992; Weil \& Hernquist 1994; Heyl, Hernquist 
\& Spergel 1996; Weil \& Hernquist 1996). From an observational
perspective, such large misalignments are rare: Franx, Illingworth 
\& de Zeeuw (1991) find much smaller misalignments for the majority 
of giant elliptical galaxies based on long-slit spectroscopy of 
their integrated light.

There are various ways to surmount some or all of these difficulties
within the context of the merger hypothesis: (1) the stellar disks may 
initially harbor compact bulges with high phase-space densities 
(Hernquist 1993); (2) the disks may contain significant amounts 
of gas (Barnes \& Hernquist 1996); or (3) ellipticals
may be the result of multiple, dissipationless mergers (Weil \& Hernquist 
1994; 1996). From a purely philosophical perspective, the first option 
is obviously unsatisfactory given that the initial conditions for spiral-spiral
mergers require a 
significant fraction of the initial mass to reside in compact, dynamically hot 
components. In the words of Hernquist (1993), ``It seems likely that even the
most strident critics of the merger hypothesis would admit the
possibility that mergers of small ellipticals will yield big
ellipticals." 

The second alternative --- that giant ellipticals form during
dissipative mergers of pairs of stellar disks, each of which contain 
significant gas components --- is closely related to the Ashman \& Zepf 
(1992) model for the formation of GCSs in giant elliptical 
galaxies. In the specific case of M87, the most compelling piece of 
evidence for a past major merger is the possible flip in the position 
angle of the rotation axis for the metal-poor GCS; we caution, however, 
that this interpretation of the observed discontinuity is not unique 
(see \S 5.7). As a further complication, the uncertainties involved in 
correctly modeling the effects of gas cooling, feedback and star formation 
in dissipational spiral-spiral simulations are formidable. To date, the 
most detailed study of merging galaxies to include gas-dynamical effects 
is that of Barnes \& Hernquist (1996), who approximated the interstellar 
medium as an incompressible fluid.
They found that the inclusion of a gaseous component has a
dramatic effect on the properties of the merger end-products,
and may alleviate some of the problems described above for the
case of purely stellar mergers: i.e., including gas leads to
merger remnants which have somewhat higher central densities, 
and may also produce closer alignments between the angular momenta and 
minor axes. These conclusions, however, are highly dependent on how
the gas is assumed to cool (see, e.g., Mihos \& Hernquist
1994, and especially \S 4.3 of Barnes \& Hernquist 1996).

Simulations of merging stellar disks have
shown that the angular momentum axes of the merger remnants often
show large misalignments with the minor axes. At large radii, metal-poor
and metal-rich GCSs in M87 both rotate around
axes which are closely aligned with the minor axis. If the
GCs are accurately tracing the kinematics of the underlying galaxy mass,
this alignment may constitute a difficulty for the major merger model;
that is to say, such a small misalignment in the 
merger remnant is by no means implausible in
the major merger model, but neither is it predicted. In their
simulations of gas-free mergers of equal-mass disks, Heyl, 
Hernquist \& Spergel (1996) find the distribution of misalignment
angles to be nearly flat, with occasional misalignments of 
75$^{\circ}$ or 80$^{\circ}$ although, as
mentioned above, gas may play a role in
ameliorating these large misalignments. 

It is clear that additional research into the influence of gas on the 
non-axisymmetry of merger remnants is needed urgently. Moreover, it is important
to quantify the expected angular momentum content and rotational properties of 
any stars or GCs which might form during dissipational mergers. In 
the case of M87, both the metal-poor and metal-rich samples show significant 
rotation, with $\langle{({\Omega R})/{{\sigma}_p}}\rangle \sim$ 0.45. This 
value is intermediate to the disparate values of 
$\langle{({\Omega R})/{{\sigma}_p}}\rangle \sim$ 1 (C\^ot\'e 1999) and 
$\langle{({\Omega R})/{{\sigma}_p}}\rangle \sim $ 0.1 (Zepf et~al. 2000)
observed for the metal-rich GCs belonging to the Milky Way and NGC 4472,
respectively. Yet the metal-rich GCs in both of these galaxies have 
been identified as the end-products of gas-rich mergers (i.e., 
Ashman \& Zepf 1992 and Zepf \& Ashman 1993 for the Milky Way; and 
Zepf et~al. 2000 for NGC 4472).

In the dissipationless hierarchical growth scenario, the simulations of
Weil \& Hernquist (1994; 1996) indicate that the rotation
axis of the merger remnant is expected to be more closely aligned with 
the photometric minor axis.\footnotemark
\footnotetext{Direct
evidence that this process has played a role
in the formation and evolution of M87 has been presented by Weil,
Bland-Hawthorn \& Malin (1997) who identified a diffuse plume
of stellar material at large radii --- the likely result of the
accretion of a small satellite galaxy.} Strictly
speaking, these simulations --- which do not include gas --- apply to merger
remnants of small
virialized groups of galaxies, although the generic kinematic properties
of the end-products are similar to those found in cosmological 
simulations of galaxy formation through the hierarchical
agglomeration of smaller galaxies and protogalactic fragments
(Barnes \& Efstathiou 1987; Frenk et al. 1988; Quinn \& Zurek 1988;
Warren et al. 1992; c.f. Dubinski 1998). In general, the assumed 
initial conditions would be appropriate if centrally dominant galaxies 
such as M87 are the result of mergers of small subgroups which later
coalesced into larger structures, as originally suggested by White (1982).
The end-products of these simulations appear
nearly round in most projections, and exhibit small kinematic 
misalignments when integrated over the full range in radius. Near their
centers, 
they show relatively little rotation, but can rotate rapidly in their outer
regions,
with $0 \la v_{\rm rot}/{\sigma} \la 0.8$, although this conclusion is likely
to depend on the assumed orbital properties of the progenitor galaxies 
(Dubinski 1998). 

In short, these results are broadly consistent with the observed behavior of 
the metal-poor GCS in M87, but the poorly-constrained 
rotation curve at large radii makes definite conclusions impossible. There
is also some evidence that the inferred rotation of the M87 GCS (particularly 
that of the metal-poor component) may include a contribution from large-scale
streaming motions along the Virgo principal axis (see below). And while
the simulations of Weil \& Hernquist (1996) produce end-products with low 
triaxialities and roughly oblate shapes (consistent with the results of
dynamical modeling of the line-of-sight velocity profile in the inner
regions of M87; e.g., van der Marel 1991; 1994; Bender et al. 1994), it
is important to bear in mind that the N-body simulations, though
similar in some respects to 
the hierarchical growth model of C\^ot\'e et al.  (1998), are not completely
analogous. Most notably, the N-body simulations do not include a single
dominant progenitor galaxy, whereas in the agglomeration picture of C\^ot\'e
et al. (1998), the metal-rich clusters arise in the dissipative (monolithic)
collapse of this component.

A more definitive test of these scenarios will require a crucial
piece of evidence which is currently lacking: accurate age estimates for 
the metal-rich and metal-poor GCs. To date, the limited 
observational evidence have proven inconclusive.  Kundu et al. 
(1998) argued, on the basis of broadband $VI$ photometry, that the 
metal-rich GCs in M87 are 3-6 Gyr younger than their metal-poor 
counterparts, as expected in the major merger model model. However, 
this claimed age difference is roughly the same size as the 
uncertainties involved in using broadband $VI$ colors to derive GC
ages, and is in contradiction with the findings of Cohen et al. 
(1998) who, from Keck spectroscopy of the brightest clusters in M87, 
found the metal-rich and metal-poor GC populations to be both
old and coeval (i.e., $T \sim 13$ Gyr).

The principal obstacle to understanding the merger/accretion
history of M87 remains the uncertain role played by gas. From a theoretical 
point of view, improved numerical simulations that include gas-dynamical 
effects and star formation are needed urgently. On the observational front, it 
is important to measure the star formation history of the GCS {\it directly}, 
through improved age estimates for the chemically distinct GC populations
(e.g., Jordan et~al. 2002).

\begin{figure*}[b]
\vskip3.5truein
\caption{Orientation of the angular momenta axes of metal-poor
and metal-rich globular clusters surrounding M87 (blue and red arrows,
respectively) overlaid on an image from the Palomar Digital Sky Survey. A
fixed radial bin width of ${\Delta}R = 90^{{\prime}{\prime}} \simeq$ 6.5 kpc
has been adopted throughout. The length of each arrow is proportional
to the amplitude of the best-fit sine curve in each annulus.
The maximum positive rotation velocity at each radius is found 90$^{\circ}$
counter-clockwise of the direction indicated by the arrows. The image
measures $20^{\prime}\times20^{\prime}$ on a side, or 87.3$\times$87.3 kpc for our
adopted M87 distance. The luminosity-weighted mean photometric minor and major
axes for the galaxy are indicated by the solid yellow lines. The dotted yellow
curves indicate the best-fit ellipses to the surface brightness profile
from Carter and Dixon (1978). The dashed ellipse indicates the approximate
location and orientation of the galaxy's cD envelope.
\label{fig18}}
\end{figure*}

\begin{figure*}[t]
\centering \leavevmode
\epsfysize=6.0truein
\epsfbox{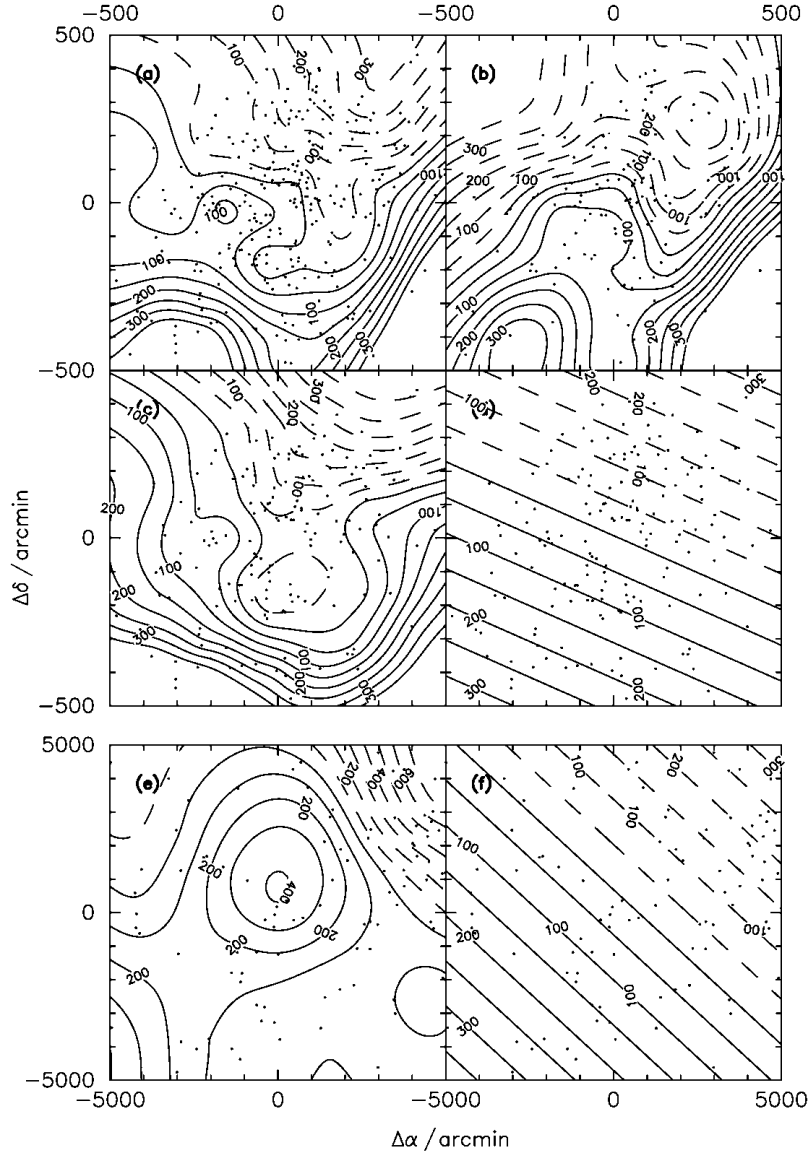}
\caption{{\it Panel a:} The line-of-sight, rotational velocity field
for the M87 globular cluster system obtained using the full sample of 278 clusters
with measured radial velocities (dots). North is up and east is to 
the left. Contours are labeled in km s$^{-1}$ with respect to the mean 
velocity. Solid contours indicate positive velocities,
while dashed contours indicate negative velocities.
{\it Panel b:} Same as (a) except for the sample of 117 metal-rich (red) globular clusters.
{\it Panel c:} Same as (a) except for the sample of 161 metal-poor (blue) globular clusters.
{\it Panel d:} Plane-of-best-fit to the rotational velocity field of metal-poor
globular clusters. Contours are labelled in km s$^{-1}$ with respect to the mean 
velocity. As in (a), solid contours indicate positive velocities, while dashed 
contours indicate negative velocities.
{\it Panel e:} The line-of-sight, rotational velocity field for 117 Virgo {\it galaxies}
located within 2$^{\circ}$ of M87 (dots). North is up and east is to 
the left. Contours are labelled in km s$^{-1}$ with 
respect to the mean velocity. As in (a), solid contours indicate positive 
velocities, while dashed contours indicate negative velocities.
{\it Panel f:} Plane-of-best-fit to the rotational velocity field of Virgo
galaxies located within 2$^{\circ}$ of M87. Contours are labelled in km s$^{-1}$ with 
respect to the mean velocity. As in (a), solid contours indicate positive 
velocities, while dashed contours indicate negative velocities.
\label{fig19}}
\end{figure*}

\subsection{Connection to the Virgo Cluster on Larger Scales}

It has long been known that the brightest of Virgo's giant elliptical
galaxies define a remarkably linear pattern on the plane of the sky 
(Arp 1968). It was recently shown that this extended arrangement, 
which passes through the cluster core at a position angle of 
110-125$^{\circ}$ and is roughly centered on M87, is the projection of a highly
collinear, three-dimensional filament --- the so-called ``principal axis'' of 
the Virgo cluster (West \& Blakeslee 2000). Thus, the rotation axis of the M87
globular cluster system not only coincides with the galaxy's minor axis at
large galactocentric radii, but is also roughly aligned with the minor axis of
the surrounding cluster. It is then natural to  ask if some fraction of the
GCS around M87 may somehow be associated with Virgo itself (White 1987; West
et~al. 1995). One mechanism that might give rise to such a population of
``intracluster'' GCs is their acquisition from other galaxies by tidal
stripping (White 1987; West et~al. 1995). Another is the possible formation
of GCs in ``failed dwarfs'' throughout Virgo: low-mass, loosely bound clouds
of gas --- analogous to, but rather more isolated than, the protogalactic
fragments thought to form stars and globular clusters in most large galaxies
(see, e.g., Harris \& Pudritz 1994; C\^ot\'e et~al. 1998, 2000) ---
that formed normal complements of (generally metal-poor) globular clusters, but were
then destroyed before much further star formation occurred, perhaps as a result
of galactic winds driven at least in part by the GC formation itself (Harris
et al. 1998; McLaughlin 2000). Whether or not the numbers of any such clouds
were enhanced in Virgo through cold-dark-matter biasing of either the GCs (West 1993)
or their host protogalactic fragments (see Ferguson \& Binggeli 1994), the plausible 
result could have been an addition
to the intracluster medium of {\it globular clusters and hot gas} that
eventually settled naturally to the bottom of the cluster's potential well,
i.e., directly around M87.

The evidence in support of such a connection between the M87 GCS and the Virgo
cluster can be summarized as follows. First, like all cD galaxies, M87
possesses an extended, low-surface-brightness envelope, and much evidence
exists that such envelopes are entities quite {\it distinct} from cD galaxy
cores. For instance, these galaxies obey the $D_{n}$-${\sigma}$ relation
defined by normal ellipticals only after the luminosity excess above an
$R^{1/4}$-law fit to the surface brightness profile of the interior is
excluded (Tonry 1987); the total luminosity of cD envelopes correlates with
several cluster properties including cluster richness and X-ray luminosity;
the surface brightness profiles of the envelopes obey density profiles that
roughly match that of the galaxies in their parent clusters (Schombert 1988);
and the ellipticity and position angle of the galaxy isophotes often show
dramatic discontinuities at the radius where the envelope begins to dominate
the surface brightness profile. This latter issue is particularly germane in
the case of M87. McLaughlin, Harris \& Hanes (1993, 1994) showed that the GCS
{\it itself} exhibits a sharp jump in surface density which coincides with the
onset of the stellar cD envelope, and that the ellipticity and orientation of
the GCS roughly matches that of the underlying envelope. Moreover, as was
described in \S 4, the so-called ``excess'' of globular clusters at large
radii in M87 is now recognized to be {\it not} a genuine surfeit, but a
consequence of the fact that the GCS-to-baryon mass ratio there has the same
value, ${\rho}_{\rm cl}(r) /[{\rho}_{\rm gas}(r)+\rho_{\rm stars}(r)] =
0.0026$, that is exhibited by nearly one hundred other galaxies (McLaughlin
1999b). Since the stellar component of M87 dies out relatively rapidly beyond
$r\ga 50$ kpc (e.g., Fig.~\ref{fig11}), the natural interpretation is that the
distant GCs are associated with the X-ray gas around M87 --- and that gas
is itself at the virial temperature of the Virgo {\it cluster} dark matter
halo (see, e.g., McLaughlin 1999a).

The dynamical analysis presented in this paper adds to this line of argument.
In particular, it is  noteworthy that the position angle of the
rotation axis of the metal-poor GCs shows an apparent discontinuity close to
the onset of the cD envelope. To aid in the visualization of this
discontinuity, Figure~\ref{fig18} shows the orientation of the angular momenta
axes of metal-poor and metal-rich GCs surrounding M87 overlaid on an image
from the Palomar Digital Sky Survey. The blue and red arrows refer to the
metal-poor and metal-rich GCs, respectively. The length of each arrow is
proportional to the amplitude $(\Omega R)$ of the best-fit sine curve in
radial bins of fixed width ${\Delta}R = 90^{{\prime}{\prime}} \simeq$ 6.5 kpc
(see Fig.~\ref{fig8}). The dotted yellow curves show the best-fit ellipses to
the surface brightness profile of the galaxy, taken from Carter and Dixon
(1978). The ellipse nearest the onset of the cD envelope is shown by the
bold, dashed curve. Recall from \S 4 that in the limit of $r \rightarrow
\infty$, $n_{\rm blue}/n_{\rm red} \rightarrow 4$. Thus, the metal-poor
component dominates the GCS at these radii, although {\it both} components
(i.e., the metal-rich {\it and} the metal-poor cluster systems) appear to
rotate around an axis which is roughly aligned with the minor axis of the
large-scale structure in Virgo. At the largest observed radii, the rotation
curve of the GCS approaches the circular velocity of Virgo itself (although
the uncertainties in the fitted values of ${{\Omega}R}$ are large at these
distances).

To examine the apparent rotation of the M87 GCS in more detail, and to
explore the possible connections with bulk motions in the Virgo cluster, we
have generated two-dimensional maps of the GC velocity fields using the
non-parametric techniques described in Gebhardt et al. (1995) and
Merritt, Meylan \& Mayor (1997). In this
approach, the cluster-by-cluster velocity residuals about a mean,
line-of-sight velocity are smoothed with a ``thin-plate smoothing spline" 
which is chosen on the basis of a generalized cross validation technique 
(Wahba 1990).  The results of this exercise are presented in Figure~\ref{fig19}.
Panels (a-c) in this figure show smoothed velocity fields for the full sample
of GCs, the metal-rich subsample and the metal-poor subsample, respectively.
A striking feature of this figure (panel b) is the form of the rotational
field for the metal-rich GCs: i.e., it appears non-cylindrical in nature, 
consistent with the discussion around equations (\ref{eq3})-(\ref{eq5}) in
\S 3. Rather, the rotation field shows
a ``double-lobed" pattern with maxima at $R \sim$ 3.5-4$R_e$ (25-30 kpc) along 
the approximate photometric major axis of the galaxy, although additional
radial velocities may be needed to fully characterize the true two-dimensional
rotation field.  Panel (c) suggests that
the rotation field of the metal-poor GCS may more closely
resemble that of a solid body; or, alternatively, it may
show evidence for a ``shear'' in the line-of-sight velocity. The magnitude and 
orientation of this putative shear is apparent in panel (d) of this Figure,
which shows the plane of best-fit to the velocity residuals for the metal-poor
GCS. Interestingly, the sharp flip in the rotation axis of the metal-poor 
clusters seen in the previous Figure is not obvious in this (smoothed) representation
of the two-dimensional velocity field, although there does appear to be a 
local minimum in the rotation velocity due south of the galaxy's center at a 
distance of $R \sim 1.5R_e$ (7 kpc).

Based on the kinematics of dwarf galaxies and X-ray imaging, Binggeli (1999)
has argued that the Virgo cluster, and particularly its dwarf galaxies, are 
not yet in dynamical equilibrium. Rather, material appears to be infalling onto
M87 along the principal axis of the Virgo cluster. It is notable that this
axis passes though both M87 and M86 --- each of these supergiant elliptical
galaxies is embedded in a swarm of dwarf elliptical galaxies and thereby defines
a major Virgo subcluster. It is therefore natural to ask if the kinematics of 
the GCS surrounding M87 is related to that of the more distant Virgo
{\it galaxies}. Such a comparison is especially relevant to the notion that
some fraction of the M87 GCS might be composed of the accreted and/or infalling
remains of ``failed dwarfs''.

\begin{figure*}[t]
\centering \leavevmode
\epsfysize=4.0truein
\epsfbox{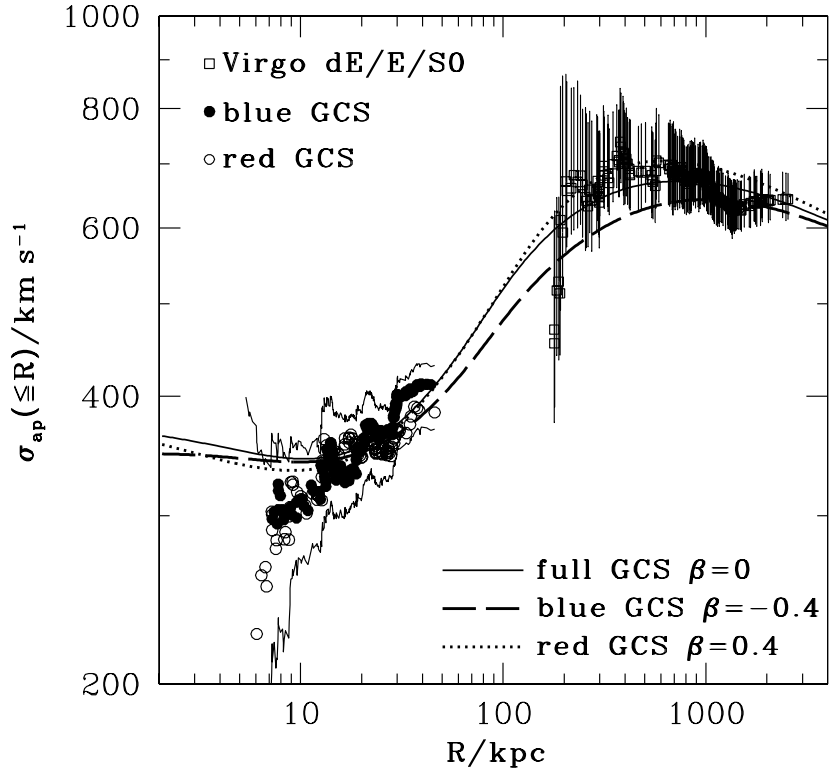}
\caption{Aperture dispersion profiles of the metal-poor (blue) and 
metal-rich (red) globular clusters around M87, and of the early-type galaxies in the Virgo Cluster at
large radii (data from Girardi et al.~1996, shown with 95\% confidence limits as
error bars). Curves are the full extent of the models for the blue globular cluster system with a
fixed orbital anisotropy of $\beta\equiv -0.4$ (cf.~Figures~\ref{fig14}), for the
red globular cluster system with $\beta\equiv+0.4$ (cf.~Figure~\ref{fig15}), and for the two
subsamples combined with a fixed $\beta\equiv0$ (cf.~Figure~\ref{fig13}).
\label{fig20}}
\end{figure*}

In order to compare the large-scale kinematics of Virgo directly to that of
the M87 GCS, we have carried out the same smoothing procedure described above
for the Virgo cluster galaxies. Using the NASA Extragalactic Database, we
identified a sample of 117 galaxies which are located within 2$^{\circ}$ 
(525 kpc) of M87 and have measured radial velocities in the range
$-1000 \le cz \le 3500$ km s$^{-1}$. The two-dimensional rotation field for
this sample is shown in panel (e) of Figure~\ref{fig19}; the corresponding
plane of best-fit to the observed velocity residuals is shown in panel (f).
Although the limited sample size makes firm conclusions impossible, it does
appear that the Virgo galaxies show evidence for a gradient in their
line-of-sight velocity which is roughly in the same direction as that observed
for the distant metal-poor GCs surrounding M87 --- broadly consistent with the 
infall picture.\footnotemark\footnotetext{Note that the observed ``shear'' is 
probably {\it not} due to the projection of the Virgo space velocity onto
the plane of the sky. The expected magnitude of this effect is 
$v_{\rm proj} = v_{\rm s}{\sin}{\Omega}$, where $v_{\rm s}$ is the cluster 
space velocity and $\Omega \simeq 0.3^{\circ}$ is the angle spanned by the GCs 
with measured velocities. Thus, for an assumed Virgo space velocity of 
$v_{\rm s} = 2000$ km s$^{-1}$, $v_{\rm proj} \sim$ 10 km s$^{-1}$, {\it far}
less than the size of the velocity gradient seen in panel (d) of
Figure~\ref{fig19}.}

Finally, we consider the {\it one-dimensional} velocity dispersion profile of
the M87 GCS in the larger context of the surrounding Virgo cluster.
Figure~\ref{fig20} shows the measured and predicted GC aperture dispersion
profiles from Figures~\ref{fig13}-\ref{fig15}, placed now on an
expanded radial scale that also includes data for 246 early-type (mostly
dE) Virgo {\it galaxies} from Girardi et~al. (1996). Error bars on the latter
measurements refer to 95\% confidence limits. The three curves show the
predicted $\sigma_{\rm ap}$ profiles corresponding to the GC subsystems and
orbital anisotropies indicated in the lower right corner of the figure.
Cases of purely radial orbits, ${\beta}$ = 1, are not shown since all such
models are ruled out at high confidence
(see Figures~\ref{fig13}--\ref{fig15}).\footnotemark\footnotetext{Indeed,
the fact that purely radial orbits are so strongly ruled out
for the GCS {\it as a whole} suggests that any infall component of GCs on
high-eccentricity orbits probably constitutes only a {\it subsample} of
the overall system.} Given that the early-type galaxies in Virgo trace the
cluster's dark-matter density profile (McLaughlin 1999a), the striking match
seen here between the kinematics of the GCs and the galaxies lends support
to our earlier assertion (recall eqs.~[\ref{eq12}]--[\ref{eq14}] and
Fig.~\ref{fig12}) that, in the limit of large galactocentric radius, 
the M87 GCS also traces the Virgo dark matter. Thus, the properties of the globular
clusters surrounding M87 are consistent with their being an ensemble of test
particles --- identical in this regard to the Virgo {\it galaxies} --- which
happen to be orbiting in the potential well defined by {\it both} the galaxy
{\it and} its parent cluster.

\placefigure{fig18}

\placefigure{fig19}

\placefigure{fig20}

\section{Summary}

We have presented a dynamical analysis of the GC system surrounding M87, the
cD galaxy at the dynamical center of the Virgo cluster.  Our database consists
of new wide-field imaging in the Washington C and T$_1$ filters, and radial
velocities for 278 GCs taken from the literature or obtained during a 
dedicated spectroscopic survey conducted with the Canada-France-Hawaii telescope. This 
constitutes the largest sample of radial velocities for pure Population II 
tracers yet assembled for any galaxy. A complete description
of the database is presented in a companion paper (Hanes et al. 2001).

Surface density profiles measured from our CT$_1$ images reveals the 
metal-poor GCS to be more spatially extended than the metal-rich population,
consistent with earlier findings based on
{\sl HST} imaging in the central regions of the galaxy. 
Beyond a radius of $R \simeq 1.5R_e$ (10 kpc), the metal-poor clusters
dominate the total GCS. The combined, metal-poor and metal-rich samples 
all show significant rotation of mean amplitude ${\Omega}R \simeq 170$ km s$^{-1}$ 
about axes whose position angles are indistinguishable from that of the 
photometric minor axis, $\Theta_0 = 65^{\circ}$. The metal-rich GCS shows
a roughly flat rotation curve of mean amplitude $\Omega R = 160_{-99}^{+130}$ km s$^{-1}$. 
Apart from a possible drift in position angle of the rotation axis at
$R \sim$ 2$R_e$, the metal-rich clusters appear to be rotating everywhere
about the photometric minor axis of the galaxy. The rotation field, however, 
appears non-cylindrical --- a two-dimensional map of the
rotation field for the metal-rich GCs shows a ``double-lobed" pattern,
with maxima at $R \sim$ 3.5-4$R_e$ (25-30 kpc) along the approximate
photometric major axis of the galaxy.

While the mean rotation velocity of the metal-poor GCS, 
$\Omega R = 172_{-108}^{+51}$ km s$^{-1}$,
is indistinguishable from that of its metal-rich counterpart, a two-dimensional
map of the rotation field suggests that this system may rotate as a solid-body or,
alternatively, it may exhibit a ``shear" in the line-of-sight velocity. This
putative shear is similar to that observed for Virgo {\it galaxies} within two degrees 
of M87 --- broadly consistent with a scenario in which some, or all, of the distant,
metal-poor GCs are gradually infalling onto M87 along Virgo's principal axis.
Inside $R \simeq 2.5R_e$ (18 kpc), the approximate onset of the galaxy's cD envelope,
the metal-poor GCS appears to rotate around the photometric {\sl major} axis.

The surface brightness profile of M87 has been combined with ROSAT 
observations of the surrounding X-ray gas to construct a 
mass model for the Virgo cluster. A comparison of the observed and 
predicted GC velocity dispersion profiles based on this model, and
on the measured GC density profiles, suggests that the velocity 
ellipsoid of the composite GCS is almost perfectly isotropic. 
Dividing the sample on the basis of metallicity reveals
the metal-poor clusters to be slightly biased to tangential orbits, 
with ${\beta}_{\rm cl} \simeq -0.4$, while the metal-rich cluster 
system shows a radial bias of the roughly the same magnitude:
${\beta}_{\rm cl} \simeq +0.4$. We note that a simple mass model for 
M87 and Virgo provides a remarkably accurate match to the observed 
GC kinematics, suggesting that the GCs, as a whole, belong to both M87 
and the surrounding Virgo cluster. More detailed conclusions 
concerning the formation of M87 and its GCS must await the measurement
of precise {\it ages} for the metal-poor and metal-rich GC systems 
--- the most outstanding observational constraints bearing on this issue.

\acknowledgments
 
PC gratefully acknowledges support provided by the Sherman M. Fairchild
Foundation during the course of this work. DEM acknowledges support from NASA 
through grant number HF-1097.01-97A awarded by the Space Telescope Science 
Institute, which is operated by the Association of Universities for Research 
in Astronomy, Inc., for NASA under contract NAS5-26555. DG acknowledges 
financial support for this project received from CONICYT through Fondecyt 
grant 8000002, and by the Universidad de Concepcion through research grant 
No. 99.011.025-1.0.  The research of DAH and GLHH is supported through 
grants from the Natural Sciences and Engineering Research Council of Canada. 
DAH is pleased to thank the Directors of the Dominion Astrophysical 
Observatory and the Anglo-Australian Observatory for their hospitality 
and support during a research sabbatical. This research has made use of 
the NASA/IPAC Extragalactic Database (NED) which is operated by the Jet 
Propulsion Laboratory, California Institute of Technology, under contract 
with the National Aeronautics and Space Administration.

\clearpage

\end{document}